\documentclass[fleqn,usenatbib]{mnras}

\usepackage{newtxtext,newtxmath}

\usepackage[T1]{fontenc}

\DeclareRobustCommand{\VAN}[3]{#2}
\let\VANthebibliography\thebibliography
\def\thebibliography{\DeclareRobustCommand{\VAN}[3]{##3}\VANthebibliography}

\usepackage{graphicx}	
\usepackage{amsmath}	
\usepackage{hyperref}   
\usepackage{threeparttable}
\usepackage{booktabs}

\title[PNG from LSS 2pcf and 3pcf]{Constraining primordial non-Gaussianity from the large scale structure two-point and three-point correlation functions}

\author[Z.~Brown et al.]{Z.~Brown$^{1}$\thanks{E-mail: zbrown5@ur.rochester.edu},
    R.~Demina$^{1}$,
    A.~G.~Adame$^{2}$,
    S.~Avila$^{3}$,
    E.~Chaussidon$^{4}$,
    S.~Yuan$^{4}$,
    V.~Gonzalez-Perez$^{2,5}$,
    \newauthor
    J.~Garc\'ia-Bellido$^{2}$,
    J.~Aguilar$^{4}$,
    S.~Ahlen$^{6}$,
    R.~Blum$^{7}$,
    D.~Brooks$^{8}$,
    T.~Claybaugh$^{4}$,
    S.~Cole$^{9}$,
    \newauthor
    A.~de la Macorra$^{10}$,
    B.~Dey$^{11}$,
    P.~Doel$^{8}$,
    K.~Fanning$^{12,13}$,
    J.~E.~Forero-Romero$^{14,15}$,
    E.~Gazta\~naga$^{16,17,18}$,
    \newauthor
    S.~Gontcho A Gontcho$^{4}$,
    K.~Honscheid$^{19,20,21}$,
    C.~Howlett$^{22}$,
    S.~Juneau$^{7}$,
    R.~Kehoe$^{23}$,
    T.~Kisner$^{4}$,
    \newauthor
    M.~Landriau$^{4}$,
    L.~Le~Guillou$^{24}$,
    M.~Manera$^{3,25}$,
    R.~Miquel$^{3,26}$,
    E.~Mueller$^{27}$,
    A.~Mu\~noz-Guti\`{e}rrez$^{10}$,
    \newauthor
    A.~D.~Myers$^{28}$,
    J.~Nie$^{29}$,
    G.~Niz$^{30,31}$,
    N.~Palanque-Delabrouille$^{4,32}$,
    C.~Poppett $^{4,37,38}$,
    M.~Rezaie$^{35}$,
    \newauthor
    G.~Rossi$^{36}$,
    E.~Sanchez$^{37}$,
    E. Schlafly$^{38}$,
    D.~Schlegel$^{4}$,
    M.~Schubnell$^{39,40}$,
    J.~H.~Silber$^{4}$,
    D.~Sprayberry$^{7}$,
    \newauthor
    G.~Tarl\'{e}$^{40}$,
    M.~Vargas-Maga\~na$^{10}$,
    B.~A.~Weaver$^{7}$,
    Z.~Zhou$^{29}$, and
    H.~Zou$^{29}$
\\
\\
The authors’ affiliations are listed in Appendix~\ref{app:affiliations}.
}

\date{Accepted XXX. Received YYY; in original form ZZZ}

\pubyear{2023}

\begin{document}
\label{firstpage}
\pagerange{\pageref{firstpage}--\pageref{lastpage}}
\maketitle

\begin{abstract}
Surveys of cosmological large-scale structure (LSS) are sensitive to the presence of local primordial non-Gaussianity (PNG), and may be used to constrain models of inflation. Local PNG, characterized by $f_{\mathrm{NL}}$, the amplitude of the quadratic correction to the potential of a Gaussian random field, is traditionally measured from LSS two-point and three-point clustering via the power spectrum and bi-spectrum. We propose a framework to measure $f_{\mathrm{NL}}$ using the configuration space two-point correlation function (2pcf) monopole and three-point correlation function (3pcf) monopole of survey tracers. Our model estimates the effect of the scale-dependent bias induced by the presence of PNG on the 2pcf and 3pcf from the clustering of simulated dark matter halos. We describe how this effect may be scaled to an arbitrary tracer of the cosmological matter density. The 2pcf and 3pcf of this tracer are measured to constrain the value of $f_{\mathrm{NL}}$. Using simulations of luminous red galaxies observed by the Dark Energy Spectroscopic Instrument (DESI), we demonstrate the accuracy and constraining power of our model, and forecast the ability to constrain $f_{\mathrm{NL}}$ to a precision of $\sigma_{f_{\mathrm{NL}}} \approx 22$ with one year of DESI survey data.
\end{abstract}




\section{Introduction}
\label{sec:intro}

The fields driving (and/or present during) inflation leave imprints on primordial density fluctuations. In the case of a single scalar field, primordial deviations from average density are expected to be  distributed according to a Gaussian of mean zero \citep{maldacena2003non, bartolo2004non}, while  more complex multi-field models \citep{creminelli2011not} are expected to deviate from Gaussianity. Thus, measurements of primordial non-Gaussianity (PNG) can distinguish between different inflationary models.

For local non-Gaussianity, which depends only on the local value of the potential, the primordial potential is written as
\begin{equation}
\Phi= \phi+f_{\mathrm{NL}}(\phi^2- \langle \phi \rangle ^2) \ ,
\label{eq:fnl}
\end{equation}
where $\phi$ denotes the potential of a Gaussian random field, and $f_{\mathrm{NL}}$ is the amplitude of the quadratic correction to this potential \citep{gangui1993three,mueller2019optimizing}.

A variety of methods have been employed to measure $f_{\mathrm{NL}}$ from both the early and late-time clustering properties of cosmological matter tracers. Initially, $f_{\mathrm{NL}}$ was measured from the angular clustering of the cosmic microwave background \citep[CMB;][]{komatsu2001acoustic,komatsu2002cosmic}. At present, the bi-spectrum of CMB anisotropies observed by the Planck satellite \citep{collaboration2020planck} offers the tightest constraints, where the most probable value and 68\% confidence limits (CL) are $f_{\mathrm{NL}} = -0.9 \pm 5.1$. Studies of cosmological large-scale structure (LSS) are also promising tests of PNG. Specifically, \cite{cunha2010primordial} and \cite{hotchkiss2011quantifying} have shown that the number density of galaxy clusters is sensitive to $f_{\mathrm{NL}}$. Beyond  the number density, or one-point statistics, additional sensitivity to PNG has been observed in two-point~\citep{mcdonald2008primordial,dalal2008imprints,slosar2008constraints,alvarez2014testing,biagetti2019hunt,mueller2021clustering} and three-point statistics~\citep{d2022limits,cabass2022constraints}, as characterized by the power spectrum and bi-spectrum, which are the Fourier transforms of the configuration space two-point and three-point correlation functions (2pcf and 3pcf, respectively). It was demonstrated that a simultaneous study of the power spectrum and bi-spectrum of galaxy distributions increases the sensitivity to PNG \citep{gualdi2021joint}. 

Analyses of LSS clustering, e.g. involving the detection of the baryon acoustic oscillations (BAO) signal, are typically carried out in both configuration and Fourier space \citep{alam2017clustering}. Here, we propose an analogous strategy for the measurement of local PNG. In the following study, we develop a method to measure $f_{\mathrm{NL}}$ by simultaneously using the configuration-space 2pcf and 3pcf. 

The effect of PNG on the clustering of cosmological matter tracers is a scale-dependent bias, where the largest deviations in clustering from the Gaussian case occur at large scales. Hence, greater sensitivity is expected from current and future large-volume spectroscopic surveys, e.g. the Dark Energy Spectroscopic Instrument (DESI). DESI is a spectroscopic surveyor attached to the Mayall 4-meter telescope at Kitt Peak National Observatory \citep{2022AJ....164..207A,2023arXiv230606310M,2023AJ....165....9S,2016arXiv161100037D}. Over an estimated five year period, DESI will measure spectra of 40 million galaxies and quasars \citep{2016arXiv161100036D}. Although the primary goal of DESI is to investigate the nature of dark energy through the Universe's expansion history \citep{2013arXiv1308.0847L}, its large volume (about a third of the sky) also makes it ideal for measuring PNG from LSS clustering. DESI will record spectra for a variety of cosmological matter tracers at different redshift ranges \citep{2020RNAAS...4..188A,2020RNAAS...4..187R,2020RNAAS...4..181Z,2020RNAAS...4..180R,2020RNAAS...4..179Y,2023ApJ...943...68L,2023AJ....165..124A,2022arXiv220808514C,2022arXiv220808512H,2023AJ....165...58Z,2023AJ....165..126R,2023ApJ...944..107C}. In this analysis, however, we study the sensitivity achievable with DESI using luminous red galaxies (LRGs) as an example tracer.

We begin by briefly introducing the algorithm used to evaluate the correlation functions, {\tt ConKer}, in Section~\ref{sec:conker_details}. We describe the method to extract the constraints on $f_{\mathrm{NL}}$ using the 2pcf and 3pcf in configuration-space in Section~\ref{sec:method}. In Section \ref{sec:desi_sens} we discuss the application of this method to simulated DESI LRGs and present forecasts of DESI sensitivity to PNG. We discuss the implications and challenges of this method in Section~\ref{sec:discussion} and conclude in Section~\ref{sec:conclusions}. 


\section{Evaluation of 2pcf and 3pcf using ConKer} 
\label{sec:conker_details}

We compute the 2pcf and 3pcf using the {\tt ConKer} algorithm described in \cite{brown2022conker}. Employing techniques developed in \cite{brown2021algorithm}, this algorithm estimates correlation functions by convolving the matter density field with spherical shell kernels. Here, we provide an overview of the method and introduce the cases of relevance for this study.

\begin{figure}
\includegraphics[width=\linewidth]{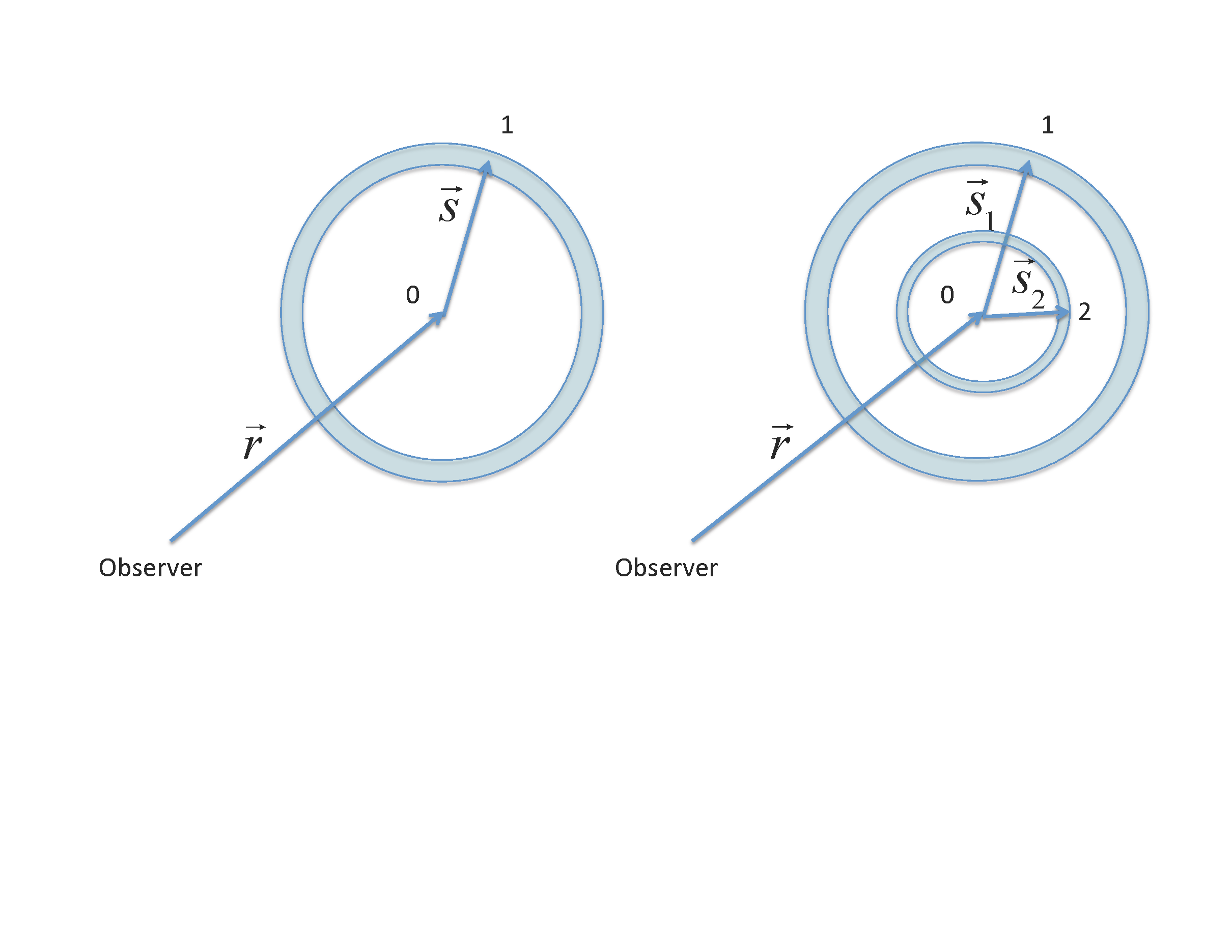}
\vspace{-2.7cm}
\caption{An illustration of the two ($0-1$) and three ($0-1-2$) point correlation function evaluation using {\tt ConKer}. The blue spherical kernels of radii $s$ for the 2pcf and $s_1$ and $s_2$ for the 3pcf are convolved with the matter distributions. The integration over $\vec{r}$ is performed by scanning the position of the kernel center $0$ over the surveyed volume.}
\label{fig:n-pletes}
\end{figure}

For an arbitrary matter tracer, let $\rho(\vec{r})$ be the tracer density at position $\vec{r}$. The expected density of a random distribution of that tracer is $\bar{\rho}(\vec{r})$. We define the deviation in density from average as 
\begin{equation}
\Delta(\vec{r}) = \rho(\vec{r})- \bar{\rho}(\vec{r})\ . 
\label{eq:delta_def}
\end{equation}
For some correlation order $n$, let us consider all possible configurations of $n$ points, referred to as  $n$-plets, with one vertex at point $0$, characterized by a vector $\vec{r}$, and the other vertices are defined by vectors $\vec{r_i}$, where $i=1,...n-1$ (see Fig.~\ref{fig:n-pletes}). Each point is connected to point $0$ by the vector $\vec{s_i}=\vec{r_i}-\vec{r}$.  The $n$-point correlation functions ($n$pcfs) characterize the excess of a particular $n$-plet over that of a random distribution of tracers. 
In this approach, the $n$pcfs are given by
\begin{equation}
\begin{split}
\xi_n(\vec{s}_1, ..\vec{s}_{(n-1)}) = \frac{1}{R_n^0} 
\int \Delta(\vec{r}) \Delta(\vec{r}_1)  ..\Delta(\vec{r}_{(n-1)}) d\vec{r}   \ , 
\end{split}
\label{eq:npcf_definition}
\end{equation}
where the integration over $\vec{r}$ implies all possible positions of point $0$ in the surveyed volume. The normalization $R_n^0$ is evaluated using the same integration performed over the random distribution of tracers. 

The angle-averaged $n$pcfs (or ``monopole'' in terms of the Legendre expansion) are defined as
\begin{equation}
\begin{split}
\xi_n^{0}(s_1, ..s_{(n-1)}) = \frac{1}{R_n^0} 
\int \Delta(\vec{r}) \Delta(\vec{r}_1)  ..\Delta(\vec{r}_{(n-1)}) d \hat{\vec{s}}_1 ..d \hat{\vec{s}}_{(n-1)}d\vec{r}   \ ,
\end{split}
\label{eq:npcf_iso}
\end{equation}
where the integration over the unit vectors $\hat{\vec{s}}_i$ implies averaging over the orientations of $ \vec{s_i}$. With {\tt ConKer}, these integrals are evaluated by convolving spherical shell kernels of radii $s_i$ with the matter density field. After averaging over angles, $n$pcfs depend only on the absolute distances $s_i$, and not on the orientation of the corresponding vectors. 

While {\tt ConKer} can be used to evaluate the Legendre multipoles of $n$pcfs of arbitrary order, in this study we consider only the monopoles of the 2pcf and 3pcf. Thus, for brevity we drop the superscript $0$ in the notation for the $n$pcfs. 

The distances $s$ are binned in $N_{\mathrm{bin}}$ bins. Then, the 2pcf can be presented as an $N_{\mathrm{bin}}$-dimensional vector, while the 3pcf is a symmetric $N_{\mathrm{bin}} \times N_{\mathrm{bin}}$-dimensional matrix. We refer to the special case where $s_1 = s_2$ as the diagonal 3pcf, because it corresponds to the elements that sit along the main diagonal of this matrix. 
We convert the part of the matrix below the diagonal into a vector of length $(N_{\mathrm{bin}}^2-N_{\mathrm{bin}})/2$. We define the observable vector $O_i$ as the concatenated vectors of the 2pcf, diagonal 3pcf, and non-diagonal 3pcf, 
\begin{equation}
O_i = [\xi_2(s),\xi_3(s_1=s_2),\xi_3(s_1>s_2)] \ .
\label{eq:xi_concat}
\end{equation}


\section{Description of the method}
\label{sec:method}
\subsection{Theoretical Considerations}
\label{sec:theory}

The relative deviation in density from the average cosmological matter density is defined as
\begin{equation}
\delta(\vec{r}) = (\rho(\vec{r})- \bar{\rho})/\bar{\rho}\ . 
\label{eq:reldelta_def}
\end{equation}
The matter overdensity $\delta_{\phi}$ associated with the primordial gravitational potential $\Phi$ is 
\begin{equation}
\delta_{\phi}=\alpha \Phi \ .
\label{eq:delta_phi}
\end{equation}
The expression for $\alpha$, a function of the wave-number $k$ associated with density perturbations on scales $s \sim k^{-1}$ and the redshift $z$ is 
\begin{equation}
\alpha(k,z)=\frac{2D(z)}{3\Omega_{m,0}}  T(k) \frac{g(z_{\mathrm{rad}})}{g(0)} (kd_H)^2\ ,
\label{eq:alpha_k}
\end{equation}
where $D(z)$ is the growth parameter normalized to 1 at $z=0$. The factor $g(z_{\mathrm{rad}}) / g(0)$ takes into account the difference of this normalization with respect to the early time matter-dominated epoch. 
The Hubble distance, $d_H = c / H_0 \approx 3000\ h^{-1}\mathrm{Mpc}$, is evaluated using the speed of light $c$ and the present day Hubble's constant $H_0$. The reduced Hubble constant is defined as $h=H_0 / (100 \ \mathrm{km/s/Mpc})$, and $\Omega_{m,0}$ is the present day relative fraction of the matter density.

The transfer function $T(k)$ is close to unity for scales $s>L_0=16(\Omega_{m,0} h^2)^{-1}\approx 3\ h^{-1}\mathrm{Mpc}$, which are typically considered in LSS clustering analyses. All terms preceding the dimensionless scale-dependent parameter $(kd_H)^2$ are also at the order of unity; hence, for typically relevant scales $s \ll d_H$, the factor $\alpha^{-1}$ can be considered small. 



Cosmological surveys do not directly observe the distribution of matter, mostly comprised of dark matter (DM). The degree to which an arbitrary tracer $t$ adheres to the underlying matter distribution is characterized by the bias $b_{t}$ \citep{desjacques2018large}. At large scales, this is described by a linear relation between the matter ($\delta$) and tracer ($\delta_t$) overdensities:
\begin{equation}
\delta_t(z)=b_{t}(z) \delta(z) \ .
\label{eq:delta_general}
\end{equation}
For Gaussian initial conditions, and at linear scales, the bias can be considered scale-independent and equal to 
\begin{equation}
b_{t}=\tilde{b}_{1t}(z) = b_{1t} + f \mu^2  \ ,
\label{eq:kaiser_term_def}
\end{equation}
where in the first term the redshift dependence of the linear bias of a given tracer is parametrized as $b_{1t}=b_{0t}/D(z)$ and the second term is the so-called Kaiser term \citep{kaiser1987clustering}, which accounts for redshift space distortions (RSD) caused primarily by the peculiar motion of the tracers. Here, $f=-d \log D(z)/d \log (1+z) $ is the logarithmic growth rate, and $\mu$ is the cosine of the angle with respect to the line of sight. Defining the linear bias $\tilde{b}_{1t}$ to include RSD via the Kaiser factor is similar to the formalism used in \cite{castorina2019redshift}. The effects of RSD on two point and three point clustering have been extensively studied \citep{slepian2017modelling}. In the monopoles of the 2pcf and 3pcf, RSD affects an overall renormalization of the linear bias. Importantly, this effect is scale independent.

In the presence of local PNG, the bias acquires scale dependence  encoded in the factor $\alpha^{-1}(z,k)$ of the form
\begin{equation}
b_{t}(z,k) = \tilde{b}_{1t}(z)+b_{\phi t}(z) f_{\mathrm{NL}} \alpha^{-1}(z,k) \ ,
\label{eq:delta_expanded}
\end{equation}
where $b_{\phi t}$ is a bias related to the gravitational potential, a.k.a. PNG bias, which we note is degenerate with $f_{\mathrm{NL}}$. A typical parameterization for $b_{\phi t}$ is given by the relation
\begin{equation}
b_{\phi t}=2\delta_c (b_{1t}-p_{t}) \ ,
\label{eq:bphi}
\end{equation}
where $\delta_c=1.69$ is the critical density of spherical collapse, and $p_{t}$ is a potentially tracer-dependent parameter describing the formation of said tracer in primordial overdensities \citep{barreira2020impact}. In LSS analyses the common assumption for galaxies is that $p_t = 1$. While we do not fix the value of $p$, we do impose a corresponding prior with 1 as the central value. 



Recently, \cite{barreira2022predictions} and \cite{lazeyras2023assembly} have found variation in $b_{\phi t}$ due to varying galaxy  or secondary halo properties like concentration. Additionally, \cite{ross2017clustering} and \cite{rezaie2021primordial} showed how observational systematics in LSS surveys (the Sloan Digital Sky Survey in this case) induce scale-dependent effects on the two-point clustering of cosmological tracers, regardless of PNG. It is therefor necessary to validate and confirm any potential detection of PNG in the most robust way possible to rule out any effects that mimic PNG.

The $n$pcfs of any tracer $\xi_n$ depend on the density variation to the $n^{\mathrm{th}}$ power and are thus approximately proportional to $b_{t}^n$. The dependence on $f_{\mathrm{NL}}$ can be expanded to the first order of the small parameter $\alpha^{-1}$, and as a result to the first order in $f_{\mathrm{NL}}$:
\begin{equation}
\xi_n(s, f_{\mathrm{NL}}) =  \xi^{b_1}_n(s) +\xi^{\mathrm{PNG}}_n(s,f_{\mathrm{NL}}) \ , 
\label{eq:xi_mod_sum}
\end{equation}
where $\xi^{b_1}_n(s)$ are the $n$pcfs of the underlying matter distribution in the absence of PNG ($f_{\mathrm{NL}}=0$) and $\xi^{\mathrm{PNG}}_n(s,f_{\mathrm{NL}})$ is the  contribution due to PNG proportional to the first power of $f_{\mathrm{NL}}$. 

The difference between $n$pcfs corresponding to non-zero and zero $f_{\mathrm{NL}}$ is linear in $f_{\mathrm{NL}}$ with coefficients  
\begin{equation}
A_n(s,b_{1t})  = (\xi_n(s,b_{1t},f_{\mathrm{NL}}) - \xi_n(s,b_{1t},0))/f_{\mathrm{NL}} \ .
\label{eq:A_fnl}
\end{equation}
We use this linearity to interpolate the value of the $n$pcfs at a given scale $s$ for arbitrary values of $f_{\mathrm{NL}}$.
In Appendix~\ref{app:linearity} we discuss tests of linearity and show the scales at which this assumption breaks down.
The coefficients $A_n$ are determined  in each bin in $s$ using DM halo simulations ($t=h$) with $f_{\mathrm{NL}}=0, 100$. 
This describes the size of the change in the scale dependent bias for a given $f_{\mathrm{NL}}$.

\subsection{Statistical model} 
\label{sec:model}
To construct a model of $n$pcfs of an arbitrary tracer, we use simulations of the tracer without PNG, referred to as the fiducial simulations, and halos from DM simulations with varied $f_{\mathrm{NL}}$. For the simulated dark matter halos, we use an approximated N-body, or ``FastPM'' technique \citep{feng2016fastpm} to generate halos in $(3\ h^{-1}\mathrm{Gpc})^3$ boxes. We refer to these simulations as FastPM-L3 (where L3 indicates the length of the simulation box in $h^{-1}\mathrm{Gpc}$). We use independent samples of the tracer of interest, referred to as the test simulations, to evaluate the performance of the model. At the final stage of the analysis the test simulation will be replaced by data.  RSD are included for all tracer simulations, but not for the FastPM-L3 halos, which is reflected in the definition of the corresponding bias parameters.  

Using the fiducial simulations, we evaluate the fiducial $n$pcfs, $\xi^{\mathrm{fid}}_n(s)$. These simulations are of the same tracer type as the test sample, and approximately match their spatial distribution.  It is possible that the linear bias in the fiducial simulations, $\tilde{b}_{1t}^{\mathrm{fid}}$, does not match the value observed in the test samples, $\tilde{b}_{1t}$. To allow for this difference, we scale $\xi^{b_1}_n(s)$ by: 
\begin{equation}
\xi^{b_{1}}_n(s) = \left( \frac{\tilde{b}_{1t}}{\tilde{b}_{1t}^{\mathrm{fid}}} \right)^n \xi^{\mathrm{fid}}_n(s) \ .
\label{eq:xi_mod_b1}
\end{equation}

To construct $\xi^{\mathrm{PNG}}_n(s,f_{\mathrm{NL}})$ we use coefficients $A_n(s, b_{1h})$ derived from the FastPM-L3 simulations. Since halos, $h$, and tracers, $t$, have different values of the linear bias --- $b_{1h}$ and $\tilde{b}_{1t}$ respectively, we scale $\xi^{\mathrm{PNG}}_n(s,f_{\mathrm{NL}})$ according to the redshift dependence of $\alpha$ given by Eq.~\ref{eq:alpha_k}. 
The PNG bias for the FastPM-L3 halos and tracers could also differ, which is reflected in the different values of parameter $p$: $p_{h}$ and $p_{t}$, respectively. Thus, $\xi^{\mathrm{PNG}}_n(s,f_{\mathrm{NL}})$ has the following form:
\begin{equation}
\xi^{\mathrm{PNG}}_n(s, f_{\mathrm{NL}}) = \left( \frac{\tilde{b}_{1t}}{b_{1h}}\right) ^n A_n(s) r_{\phi} f_{\mathrm{NL}} \ . 
\label{eq:xi_mod_PNG}
\end{equation}
The difference between the ratio of the PNG to the linear bias for the FastPM-L3 halos compared to the tracers is encoded in the factor $r_{\phi}$, where
\begin{equation}
r_{\phi} = \frac{b_{1 h}}{\tilde{b}_{1t}} \frac{b_{\phi t}}{b_{\phi h}}   = \frac{b_{1 h}}{\tilde{b}_{1t}} \frac{b_{1t} - p_t}{b_{1h} - p_h}\ . 
\label{eq:Rphi_def}
\end{equation}
We note that the linear tracer bias within the definition of $b_{\phi t}$ appears without the Kaiser term. 
The relationship between $b_{1t}$ and $\tilde{b}_{1t}$ is given by the Eq.~\ref{eq:kaiser_term_def}.
The Kaiser factor was decomposed into a Legendre basis in Eq. 2.19 of \cite{castorina2019redshift}. Following this, we relate the two linear biases for the monopole term as $\tilde{b}_{1t} \approx b_{1t} + f/3$. The logarithmic growth rate may be estimated from the redshift dependent matter density given by $f \approx \Omega_m(z)^{0.55}$. Here, the $z$-dependent relative fraction of the matter density is

\begin{equation}
\Omega_m(z)=\frac{\Omega_{m,0}(z+1)^3}{\Omega_{m,0}(z+1)^3+\Omega_{k,0}(z+1)^2+\Omega_{\Lambda,0}} \ ,
\label{eq:om_matter}
\end{equation}
which depends on the  present day matter, curvature, and cosmological constant relative densities $\Omega_{m,0}$, $\Omega_{k,0}$, and $\Omega_{\Lambda,0}$ respectively. 

The final expressions for the expected $n$pcfs have the following form:
\begin{equation}
\xi_n(s, f_{\mathrm{NL}}) =  \left( \frac{\tilde{b}_{1t}}{\tilde{b}_{1t}^{\mathrm{fid}}} \right)^n \xi^{\mathrm{fid}}_n(s) + \left( \frac{\tilde{b}_{1t}}{b_{1h}}\right) ^n A_n(s) r_{\phi} f_{\mathrm{NL}} \ . 
\label{eq:xi_mod_full}
\end{equation}
This equation provides the expected values of the observable vector $O_i$, defined in Eq.~\ref{eq:xi_concat}, which we denote as $\tilde{O_i}$. We use the observed and expected values of $O_i$ to construct a test statistic $\chi^2$ that depends on the parameters of interest (POIs): $f_{\mathrm{NL}}$ and the linear bias $b_{0t}$, and on a set of nuisance parameters $\theta$: 
\begin{equation}
\theta = [b_{0h}, b_{0t}^{\mathrm{fid}}, p_h, p_t]  \ .
\label{eq:params_fid}
\end{equation}
The nuisance parameters are constrained using Gaussian priors with central values given by a vector $\Theta$ with Gaussian widths $\sigma$.
The test statistic is then defined
\begin{equation}
\chi^2 = \sum_{ij}\left[O_i-\tilde{O}_i\right]^TC^{-1}_{ij}\left[O_j-\tilde{O}_j\right] + \sum_f \frac{(\theta_f - \Theta_f)^2}{(\sigma_f)^2}\ , 
\label{eq:chi2}
\end{equation}
where $C_{ij}$ denotes the covariance matrix. The first summation over indices $i$ and $j$ is performed over bins in $s$ and the second over nuisance parameters $\theta_f$.

$\chi^2$ is minimized to find the optimal values of the POIs in the test sample. In the following, we discuss the application of this method to the specific case of DESI luminous red galaxies, LRGs.
\section{Application to simulated sample of DESI LRGs} 
\label{sec:desi_sens}

\subsection{Simulations} 
\label{subsec:sim_details}
All simulations employed in this study are based on a flat $\Lambda$CDM cosmology. Simulations of LRGs are generated with $\Omega_{m,0} = 0.315$ and all FastPM-L3 halo simulations have $\Omega_{m,0} = 0.3089$. Due to the difference in the present day value of the matter density, the second term of Eq.~\ref{eq:xi_mod_full} is scaled by their ratio. For the fiducial simulations, we use an ensemble of 1000 mocks based on the effective Zel'dovich (EZ) approximation \citep{chuang2015ezmocks}. For each realization, northern galactic cap (NGC) and southern galactic cap (SGC) catalogs are generated according to the angular footprint of the complete DESI Year-5 LRG survey. The redshift distribution $n(z)$ is subsampled to match that of the LRGs in the survey validation catalog \citep{2023arXiv230606307D,2023arXiv230606308D}. We refer to these simulations as subsampled Year-5 DESI LRGs (SY5). Each realization consists of approximately 2.70M objects in the combined NGC and SGC. The cosmic variance observed across this ensemble of simulations is expected to be larger than the full DESI Year-5 LRG survey.

DESI will, however, be releasing a partial LRG catalog as a part of the Year-1 (Y1) data release. We also use an ensemble of 1000 EZ mocks corresponding to the Y1 angular and redshift coverage, allowing us to forecast the sensitivity to PNG achievable using only the Y1 LRG survey. Each realization of these simulations consists of approximately 3.27M objects, matching the DESI Y1 LRGs.
 
\begin{figure}
\includegraphics[width=\columnwidth]{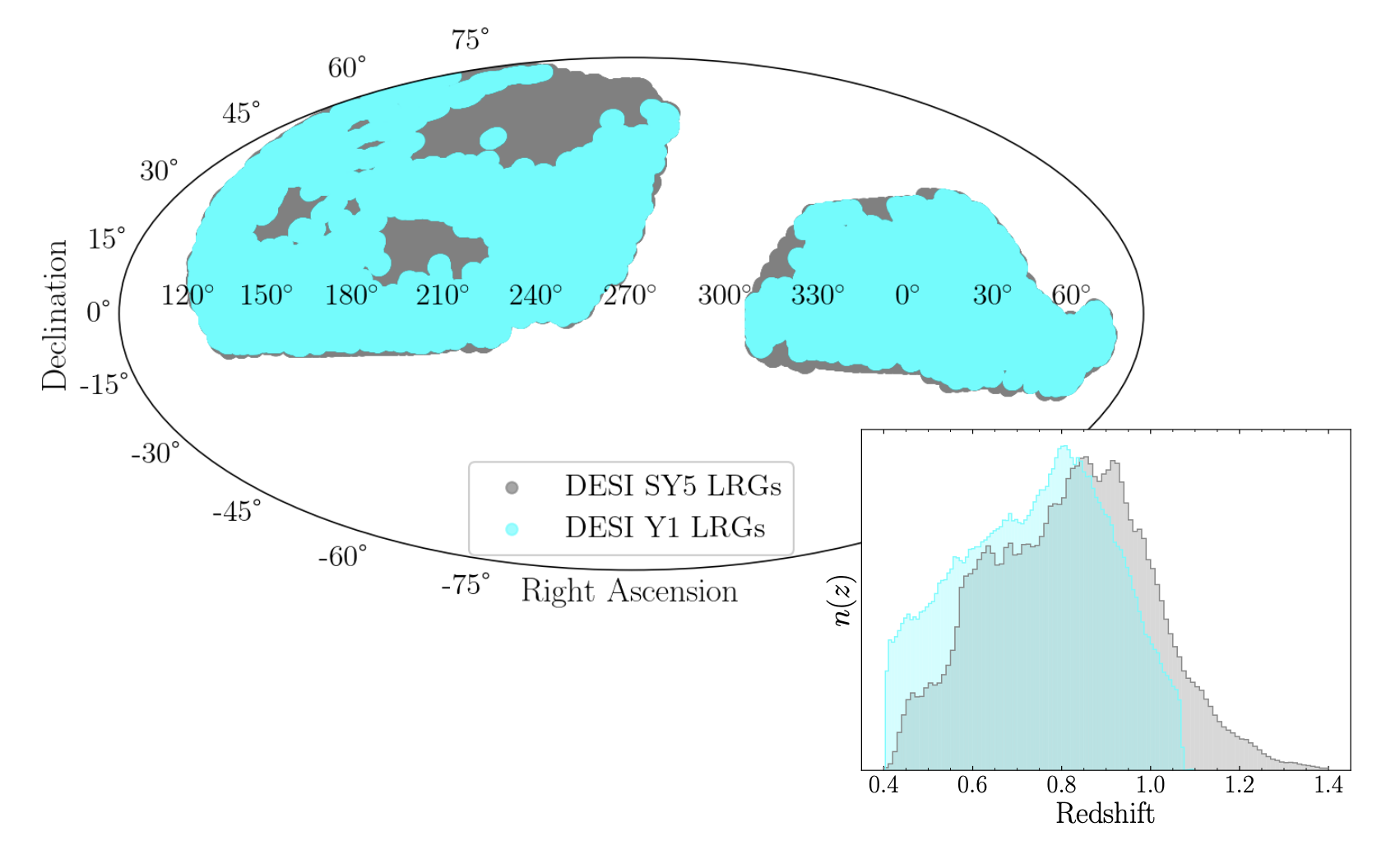}
\vspace{-0.5cm}
\caption{
The angular coverage (top left) and the redshift distribution (bottom right) of the DESI SY5 (grey) and Y1 (cyan) LRG simulations. 
}
\label{fig:desi_window}
\end{figure}

As our test samples, we consider an ensemble of higher resolution N-body simulations. They consist of 25 realizations of AbacusSummit \citep{maksimova2021abacussummit,garrison2021abacus} LRGs, distributed about halos using a halo occupation distribution (HOD) model defined in \cite{bose2022constructing,yuan2023full}. Like the SY5 EZ mocks, the AbacusSummit LRGs are convolved with the NGC and SGC DESI Year-5 angular window and subsampled to match the survey validation redshift distribution.

For both the AbacusSummit and SY5 EZ mock catalogs, we include only tracers within the redshift range $0.4 < z_g < 1.4$, where $g$ denotes galaxies. The angular coverage and redshift distributions of these simulations are shown in Fig.~\ref{fig:desi_window}.

In the FastPM-L3 halo simulations, $N=2560^3$ DM particles of mass $m_p = 1.4\times10^{11}\ h^{-1}M_{\odot}$ in a $(3\ h^{-1}\mathrm{Gpc})^3$ volume were evolved from $z=99$ to $z=1$ using 50 linearly spaced time steps. 100 realizations with $f_{\mathrm{NL}} = 0,\ 100$ were generated with matched phases, such that individual realizations at both $f_{\mathrm{NL}} = 0$ and $f_{\mathrm{NL}} = 100$ share the same initial conditions \citep{avila2023validating}. This results in two ensembles of DM halo simulations, one with and one without the presence of PNG. Catalogs were created by selecting halos with $M > 1.15\times10^{13}\ h^{-1}M_{\odot}$. Due to their size, these simulations are ideal for detecting the effects of the linear change in $n$pcfs due to PNG. Although they are likely too small to include larger modes leading to non-linear effects, our model assumes linearity in $f_{\mathrm{NL}}$ as discussed in Appendix~\ref{app:linearity}.

In addition to the FastPM-L3 halo simulations, we also use a set of four larger FastPM simulations generated in $(5.52\ h^{-1}\mathrm{Gpc})^3$ boxes. We refer to these as the FastPM-L5.52 halo simulations. Their use in this study and a description of their properties are found in Appendices \ref{app:linearity} \& \ref{app:nonzero_png}. A complete summary of all simulations used may be found in Table~\ref{tab:sim_details}.

\begin{table*}
\begin{center}
\begin{threeparttable}
\begin{tabular}{lllllll}
\toprule
Name & $f_{\mathrm{NL}}$ & Tracer & Geometry & Volume & $N_{\mathrm{real}}$ & Purpose\\
\ & \ & \ & \ & [($h^{-1}\mathrm{Gpc})^3$] & \ & \ \\
\midrule
EZ Mocks SY5 & 0 & LRGs & Subsampled DESI Y5 cut sky & 35.7 & 1000 & Fiducial model \\
EZ Mocks Y1 & 0 & LRGs & DESI Y1 cut sky & 26.3 & 1000 & Fiducial model \\
AbacusSummit & 0 & LRGs & Subsampled DESI Y5 cut sky & 35.7 & 25 & Validation tests \\
FastPM-L3 & 0, 100 & DM Halos & $L = 3\ h^{-1}\mathrm{Gpc}$ box & 27 & 100 & Estimate scale-dependant bias \\
FastPM-L5.52 & -25, 0, 12, 25 & DM Halos & $L = 5.52\ h^{-1}\mathrm{Gpc}$ box & 168.2 & 1 & Validation tests and linearity checks \\
\bottomrule
\end{tabular}
\caption{Descriptions of all simulations used, including the values of $f_{\mathrm{NL}}$, the type of tracer, the geometry of the catalogs, their volume, and their purpose in this study. $N_{\mathrm{real}}$ is the number of realizations of the simulation for each value of $f_{\mathrm{NL}}$.}
\vspace{-0.2cm}
\label{tab:sim_details}
\end{threeparttable}
\end{center}
\end{table*}


\subsection{Fiducial $n$pcfs and Covariance Matrix}
\label{subsec:fid_npcf}
\begin{figure}
\includegraphics[width=\columnwidth]{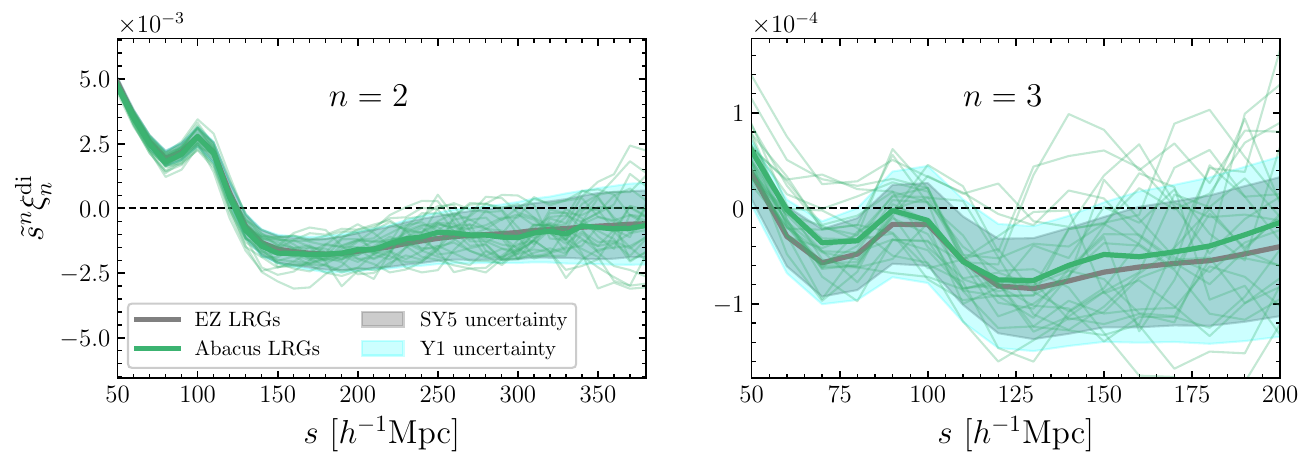}
\includegraphics[width=\columnwidth]{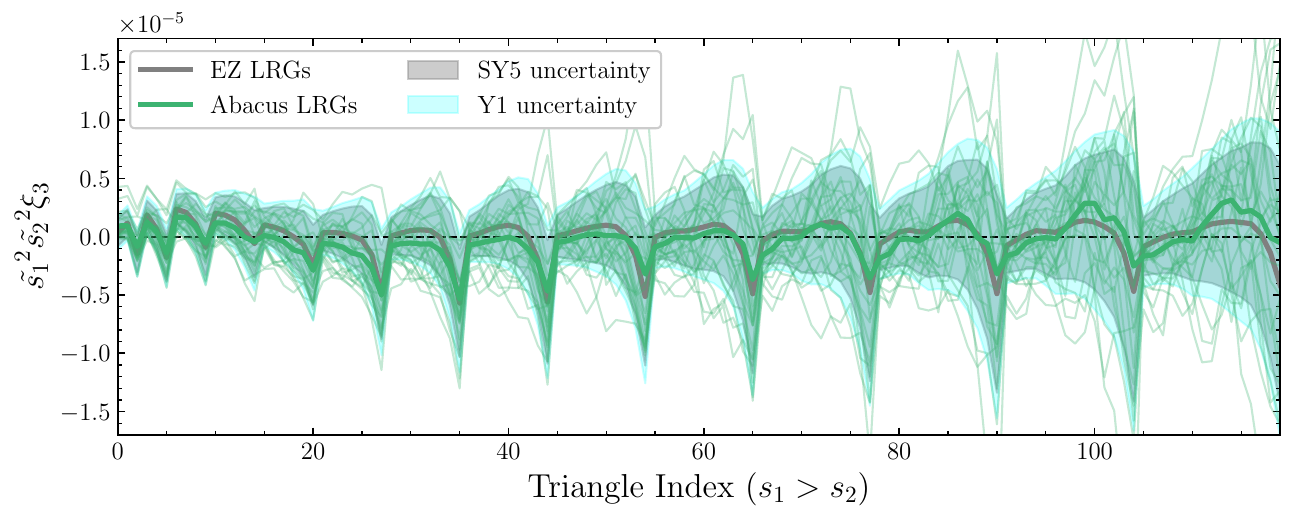}
\caption{Top: The average diagonal $n$pcfs $\xi_n^{\mathrm{di}}$ when $n=2,3$ for the SY5 EZ mocks (grey) and AbacusSummit simulation (green), as a function of the distance scale, $s$. The shaded region shows uncertainties estimated from the covariance matrix for the SY5 EZ mocks (grey) and Y1 EZ mocks (cyan). The thin green lines show the $n$pcfs for each realization of the AbacusSummit simulations independently. The diagonal $n$pcfs are multiplied by $\tilde{s}^n$, where $\tilde{s} = (s/100\ h^{-1}\mathrm{Mpc})$, to emphasize features at large scales. Bottom: The 3pcf $\xi_3$ for the SY5 EZ mocks (grey) and AbacusSummit simulation (green), as a function of the triangle index where $s_1 > s_2$. The 3pcf is  multiplied by $\tilde{s}_1^2 \tilde{s}_2^2$.  In all panels, the dashed black horizontal line denotes 0.}
\label{fig:LRG_npcf}
\end{figure}

The fiducial $n$pcfs are evaluated using the average over the ensemble of EZ mocks and are shown for the SY5 simulations in Fig.~\ref{fig:LRG_npcf}. The top panel shows the 2pcf and the diagonal part of the 3pcf. In the bottom panel of Fig.~\ref{fig:LRG_npcf}, we show the 3pcf as a function of the triangle index. For comparison we also present the results of the AbacusSummit LRGs, used as test samples. The $n$pcf uncertainties, evaluated using the covariance matrices, are also shown for the SY5 LRGs and the Y1 LRGs.


\begin{figure}
\includegraphics[width=\columnwidth]{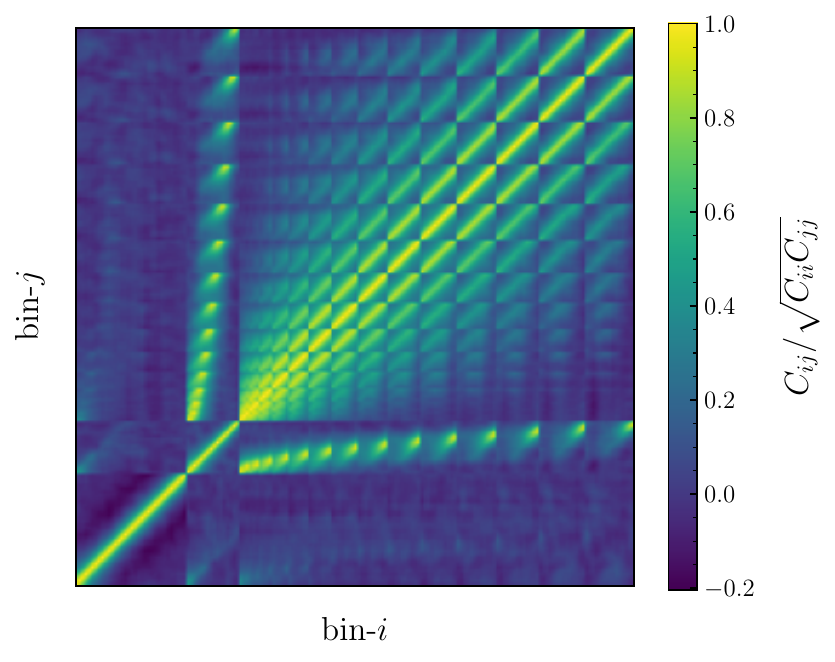}
\caption{The correlation matrix $C_{ij} = C_{ij}/ \sqrt{C_{ii} C_{jj}}$, derived from the ensemble of SY5 EZ mocks.}
\label{fig:mod_covariance}
\end{figure}

We evaluate the covariance matrix from both the ensemble of SY5 EZ mocks and from the Y1 EZ mocks. The normalized correlation matrix for SY5 is shown in Fig.~\ref{fig:mod_covariance}. Bins corresponding to the 2pcf and diagonal 3pcf are located in the bottom left of the matrix, and bins corresponding to the off-diagonal 3pcf are located in the top right.
 
\subsection{Sensitivity to PNG}
\label{subsec:png_sens}
\begin{figure}
\includegraphics[width=\columnwidth]{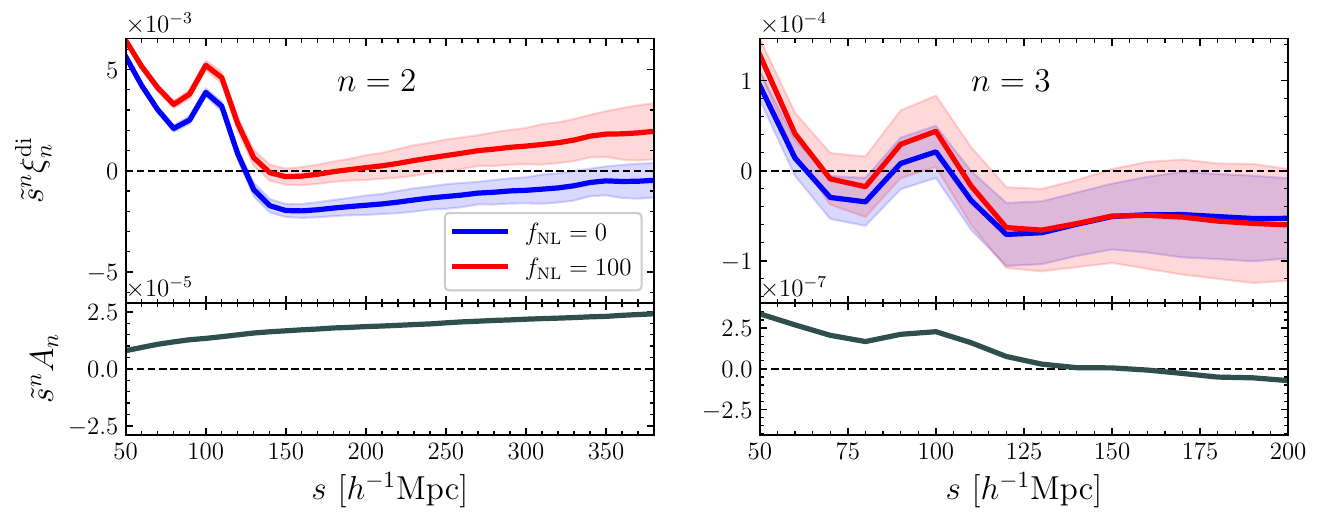}
\includegraphics[width=\columnwidth]{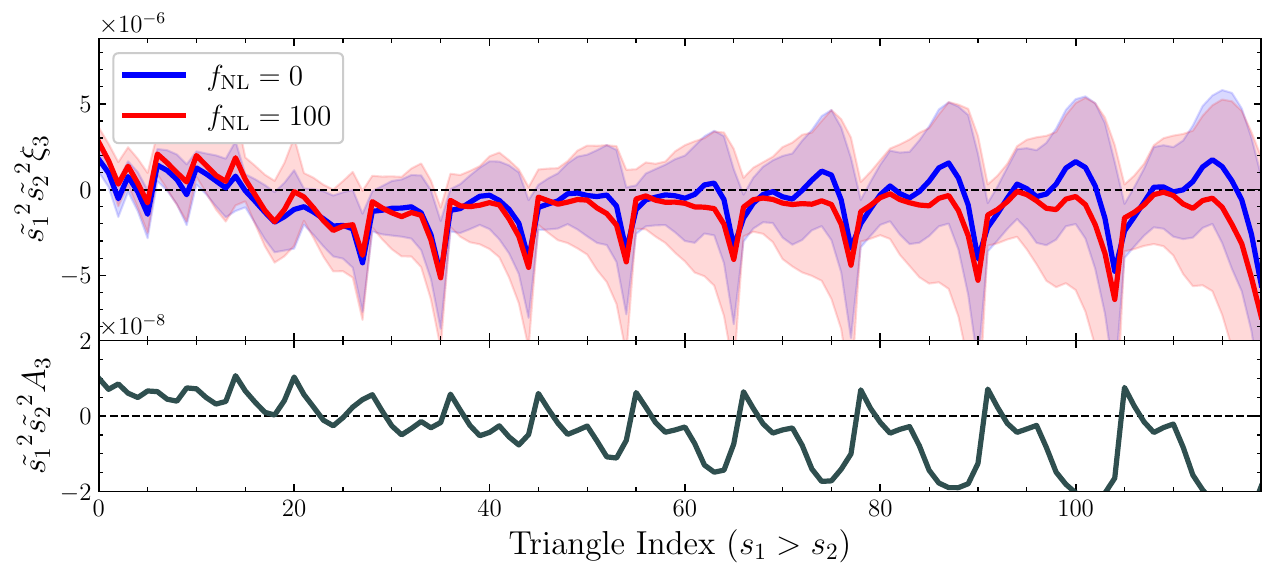}
\caption{
 The $n$pcfs $\xi_n^{\mathrm{di}}$ for $n=2,3$ (upper panels) for the ensemble of the FastPM-L3 simulated halos with $f_{\mathrm{NL}} = 0$ (blue), $100$ (red), and the estimate of $A_n$ (lower panels). The shaded regions show uncertainties estimated from the covariance matrix of each ensemble. Top: The diagonal $n$pcfs and $A_n$ multiplied by $\tilde{s}^n$ are shown as functions of the distance scale, $s$. Bottom: 3pcf and $A_3$  multiplied by $\tilde{s}_1^2 \tilde{s}_2^2$ are shown as a function of the triangle index. 
In all panels, the dashed black horizontal line denotes 0.}
\label{fig:halo_npcf}
\end{figure}

We use the FastPM-L3 halo simulations with different values of $f_{\mathrm{NL}}$ to evaluate the sensitivity of the $n$pcfs to the presence of PNG as presented in Fig.~\ref{fig:halo_npcf}. The lower panels show the coefficients $A_n$, which are computed from the deviation in the $n$pcfs for $f_{\mathrm{NL}} = 100$ from that of $f_{\mathrm{NL}} = 0$. We observe a particular sensitivity in the 2pcf where the effect on halo clustering from PNG is clearly visible. 

In all cases of the 3pcf, we find considerably less sensitivity to the presence of PNG, and large uncertainties for these monopole configurations. We do note, however, that the FastPM-L3 halos show a degree of PNG sensitivity at small scales (see the left-most portion of the $A_3$ panel). At large scales, the uncertainties are considerably larger than any deviation due to the presence of PNG. The increased statistics of the full DESI sample will reduce these uncertainties, thus enhancing the sensitivity to PNG at large scales. 


\subsection{Linear and PNG biases}
\label{subsec:lin_bias}
\begin{figure}
\includegraphics[width=\columnwidth]{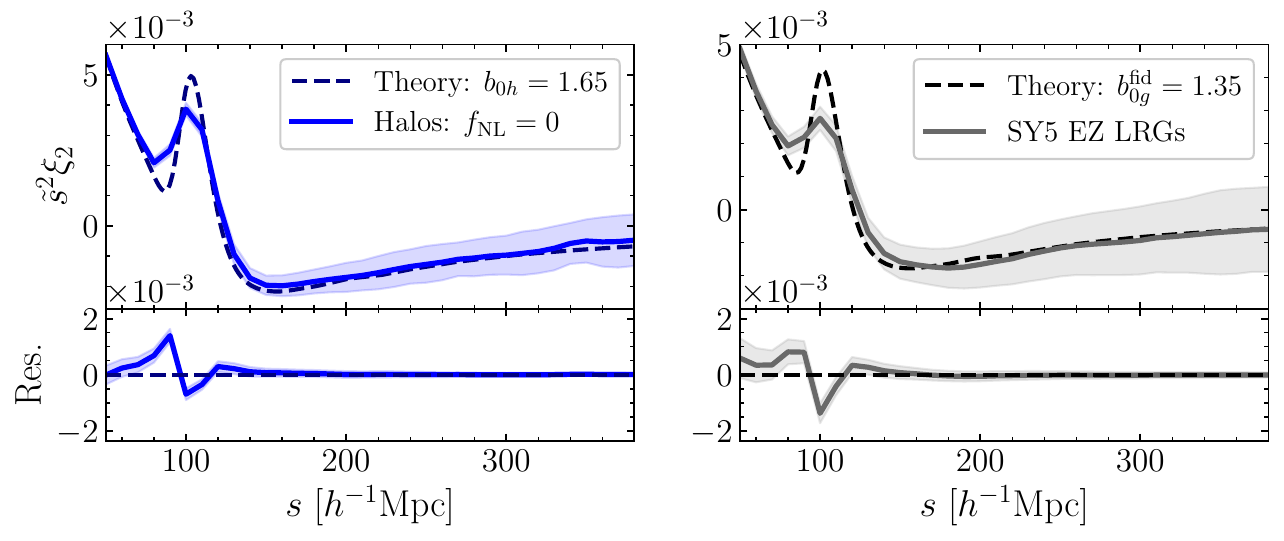}
\caption{Left: (Upper panel) The 2pcf $\xi_2$ multiplied by $\tilde{s}^2$  for the ensemble of FastPM-L3 halo simulations with $f_{\mathrm{NL}} = 0$ (solid line), the uncertainties (shaded region), and the theoretical 2pcf computed from linear theory (dashed line) scaled by the fit value of the linear bias, $[b_{0h}D(z_h)]^2$, as a function of the distance, $s$. (Bottom panel) the residual between the observed 2pcf and the scaled theoretical 2pcf. Right: The same for the ensemble of SY5 EZ mocks and the fiducial linear galaxy bias, $b_{0g}^{\mathrm{fid}}$.}
\label{fig:lin_bias_fit}
\end{figure}

The priors on the linear biases $b_{0h}$ and  $b_{0g}^{\mathrm{fid}}$ are derived using the FastPM-L3 halo and fiducial simulations respectively. To do this, we measure the ratio of the 2pcf for each tracer to the 2pcf predicted by linear theory. The theoretical 2pcf was computed using the open source cosmology toolkit {\tt nbodykit} \citep{nbodykit}. For both cases, we estimate the linear bias by performing a fit according to:
\begin{equation}
b_{1} = (\xi_{2}/\xi_2^{\mathrm{lin}})^{1/2} \ . 
\label{eq:lin_bias_fit}
\end{equation}
We show the results of this fit and the measured 2pcfs again in Fig.~\ref{fig:lin_bias_fit} for the FastPM-L3 halos and the SY5 EZ mocks. The fit is stable at the scales considered in this analysis. For the SY5 LRGs, we measure $\tilde{b}_1$ with the fit, remove the Kaiser term, and normalize to $b_{0g}$ to report the value of $b_0$ in the legend of Fig.~\ref{fig:lin_bias_fit}.

Here, $\xi_2^{\mathrm{lin}}$ is evaluated at the same redshift, or effective redshift of the sample. For the FastPM-L3 halo simulations, the entire volume has been evolved to $z_h=1$ and the effective redshift for the SY5 EZ mock LRG simulations is $z_g=0.82$. These redshifts are used to compute $D(z)$ to normalize each linear bias.


The most probable values of $b_{0h}$ and $b_{0g}^{\mathrm{fid}}$ arising from the 2pcf fits are chosen as the central values of the Gaussian priors, and we choose a width of $\sigma=0.04$ for the FastPM-L3 halos, and $\sigma=0.055$ for the LRGs, reflecting the uncertainty in our estimates. In this study we choose $1$ as the central value of the Gaussian prior on $p_h$ and $p_g$ with a width of $0.1$ for both. 


\subsection{Test samples}
\label{subsec:Toys}
To explore the projected sensitivity of the SY5 DESI LRGs, we perform two tests. First, we validate our model by measuring the parameters using the average $n$pcfs of the AbacusSummit LRGs as our observation. We treat this as an observation with the covariance matrix derived from the SY5 EZ mocks.

Secondly, we generate toy data randomly distributed around the values of the expected $n$pcfs given in Eq.~\ref{eq:xi_mod_full}, with uncertainties predicted by the covariance matrix of the SY5 EZ mocks. While the AbacusSummit simulations were generated with $f_{\mathrm{NL}} = 0$, toy experiments can be produced with arbitrary values of $f_{\mathrm{NL}}^{\mathcal{T}}$ and $b_{0g}^{\mathcal{T}}$, thus allowing us to calibrate the method on sample observations with PNG. We may also vary the values of the nuisance parameters, $\theta$, in the generation of toy model data.

\subsection{Results of the fit}
\label{subsec:Results}

We compute the 2pcf using {\tt ConKer} in $N_{\mathrm{bin}} = 34$ bins of width $\Delta s = 10$ $h^{-1}$Mpc from $s = 50\  h^{-1}\mathrm{Mpc}\ \mbox{---}\ 380$ $h^{-1}$Mpc, unless stated otherwise, for all simulation samples. The 3pcf is similarly computed in in $N_{\mathrm{bin}} = 16$ bins of width $\Delta s = 10$ $h^{-1}$Mpc from $s = 50\  h^{-1}\mathrm{Mpc}\ \mbox{---}\ 200$ $h^{-1}$Mpc. The combination of these two measurements results in a vector with $N_{\mathrm{obs}}^{\mathrm{2pcf}} + N_{\mathrm{obs}}^{\mathrm{3pcf}} = 170$ total bins. Our choices of scales are motivated by the limitation of the linearity assumption discussed in Appendix~\ref{app:linearity}.


\begin{figure*}
\includegraphics[width=\textwidth]{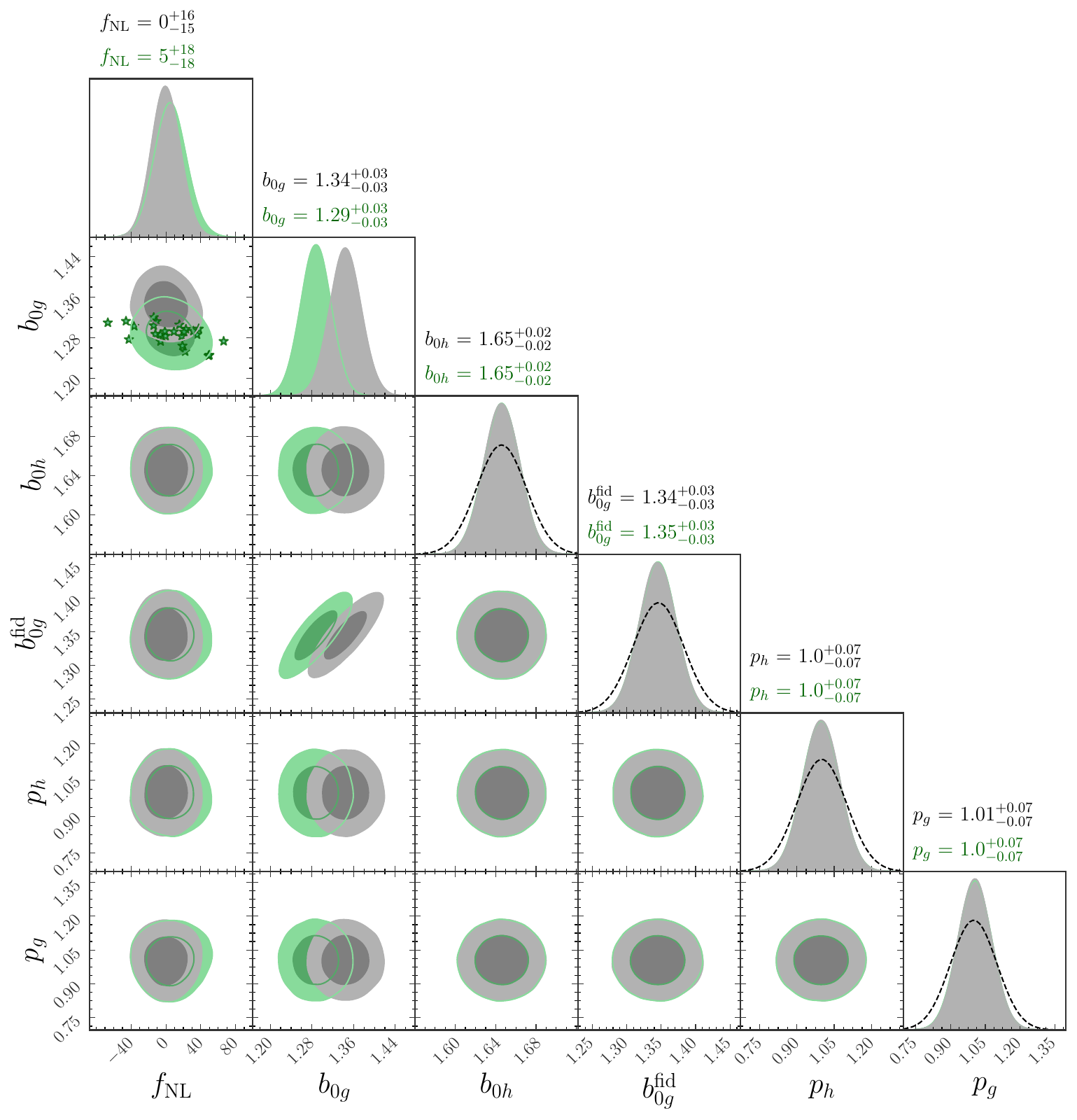}
\caption{Marginalized distributions of the POIs and nuisance parameters $\theta$, for SY5 toy model $n$pcfs (grey), and for the mean of the AbacusSummit simulations (green). The dark and light regions represent $1\sigma$ and $2\sigma$ contours, respectively. The most probable values and $1\sigma$ CL are labeled for each parameter above the respective panel. Priors are shown by dashed lines in the one-dimensional histograms. The stars in the $(f_{\mathrm{NL}}$, $b_{0g})$ panel show the most probable values for each realization of the AbacusSummit mocks.}
\label{fig:constraints}
\end{figure*}

\begin{figure}
\includegraphics[width=\columnwidth]{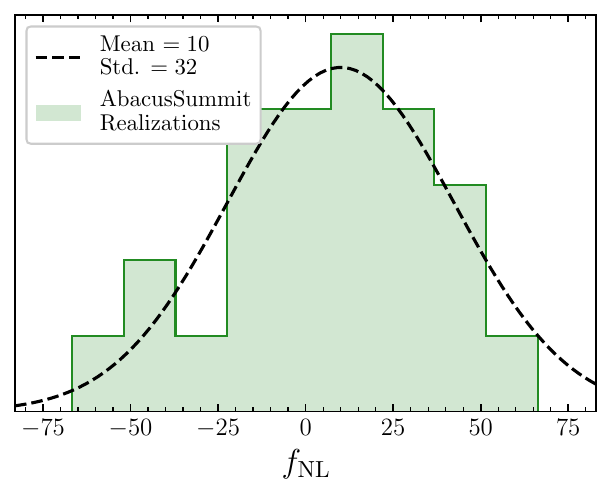}
\caption{A distribution of the most probable values of $f_{\mathrm{NL}}$ for each realization of the AbacusSummit mocks (green). The histogram bins are fit to a Gaussian (dashed line) where the width and central value of the fit are given in the legend.}
\label{fig:abacus_spread}
\end{figure}

\begin{figure*}
\includegraphics[width=\textwidth]{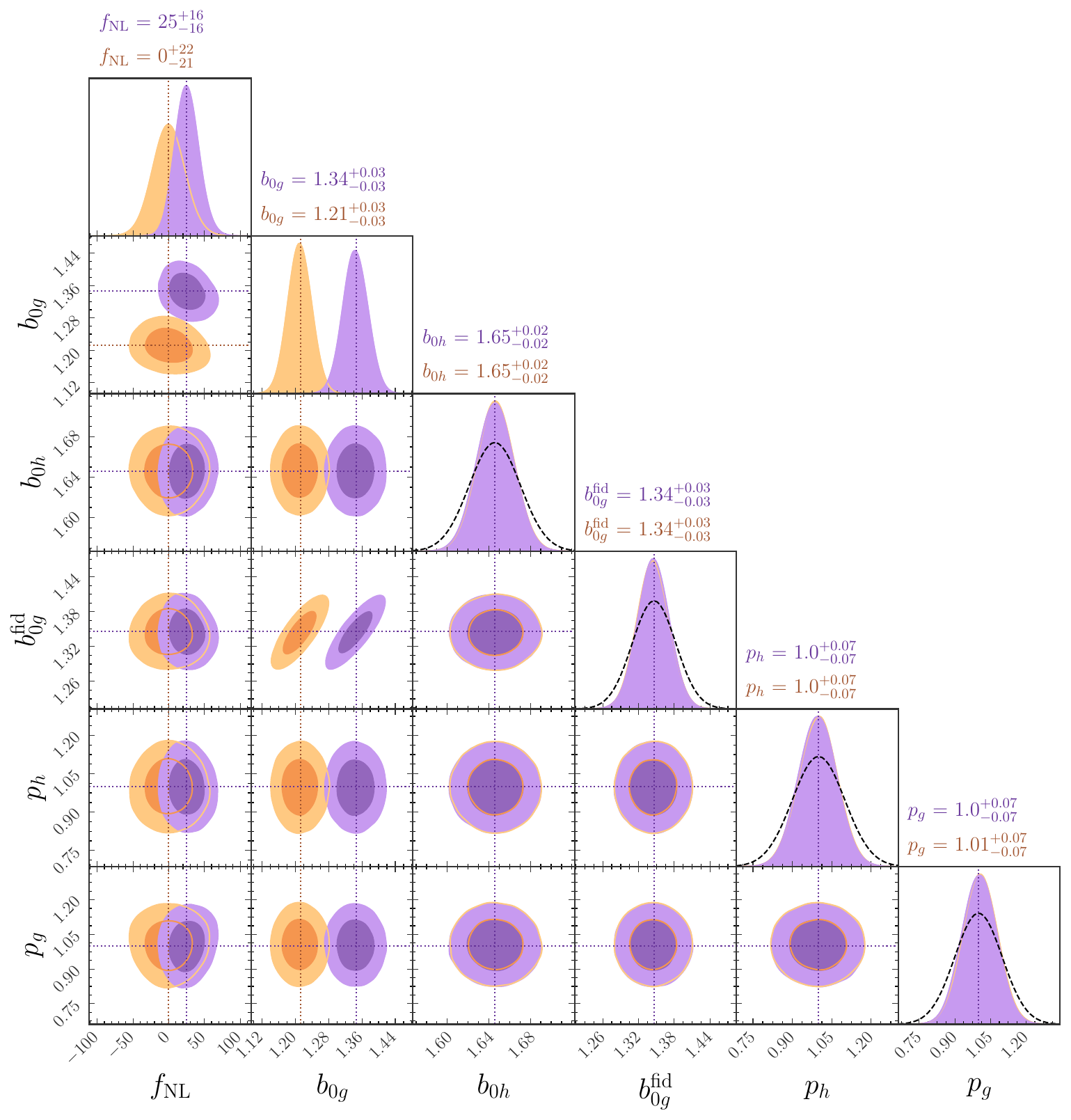}
\caption{Marginalized distributions of the POIs and nuisance parameters $\theta$, for SY5 toy model $n$pcfs with $f_{\mathrm{NL}}^{\mathcal{T}} = 25$, $b_{0g}^{\mathcal{T}} = 1.34$ (purple) and $f_{\mathrm{NL}}^{\mathcal{T}} = 0$, $b_{0g}^{\mathcal{T}} = 1.21$ (orange). The dark and light regions represent $1\sigma$ and $2\sigma$ contours, respectively. The most probable values and $1\sigma$ CL are labeled for each parameter above the respective panel. Priors are given by dashed lines in the one-dimensional histograms. The truth values indicated by the vertical and horizontal dotted lines correspond to the values of the parameters used to generate these toy model data.}
\label{fig:constraints_altPOI}
\end{figure*}

To sample the test statistic described in Eq.~\ref{eq:chi2}, we use a Markov Chain Monte-Carlo (MCMC) algorithm with 50 walkers across 10K steps, and remove the initial 500 steps of burn-in. We show the marginalized distributions over the POIs and nuisance parameters in Fig.~\ref{fig:constraints}. For the SY5 toy data, generated with $f_{\mathrm{NL}}^{\mathcal{T}} = 0$ and $b_{0g}^{\mathcal{T}} = 1.34$, the expected uncertainty on $f_{\mathrm{NL}}$ is approximately 16. 
For the validation case (the AbacusSummit simulations) we observe that $f_{\mathrm{NL}}$ is consistent with the expected value (namely $f_{\mathrm{NL}} = 0$) with a slightly larger uncertainty of approximately 18. We note that for these two cases, the value of  $b_{0g}$ differs, which can be expected for different simulations. The ability to correctly measure $f_{\mathrm{NL}}$  is retained even when the tracer bias differs from the fiducial value.

To understand the degree to which this measurement is limited by cosmic variance, we repeat the procedure (measuring the most probable values of the parameters) for each realization of the AbacusSummit mocks. We present these measurements of the POIs as stars in the $(f_{\mathrm{NL}}$, $b_{0g})$ panel of Fig.~\ref{fig:constraints}. We observe marginally larger variation in the measurements performed across each realization of the simulation, as compared to the estimated uncertainty. We show a distribution over the measured values of $f_{\mathrm{NL}}$ for each realization in Fig.~\ref{fig:abacus_spread}. The central value of the Gaussian fit to the histogram is  within the $\sigma/2$ of the true value. The variation between realizations of the AbacusSummit mocks is slightly larger than the projected $f_{\mathrm{NL}}$ uncertainty (32 vs expected 19), which is not surprising given the small number of AbacusSummit mocks.



\begin{figure}
\includegraphics[width=\columnwidth]{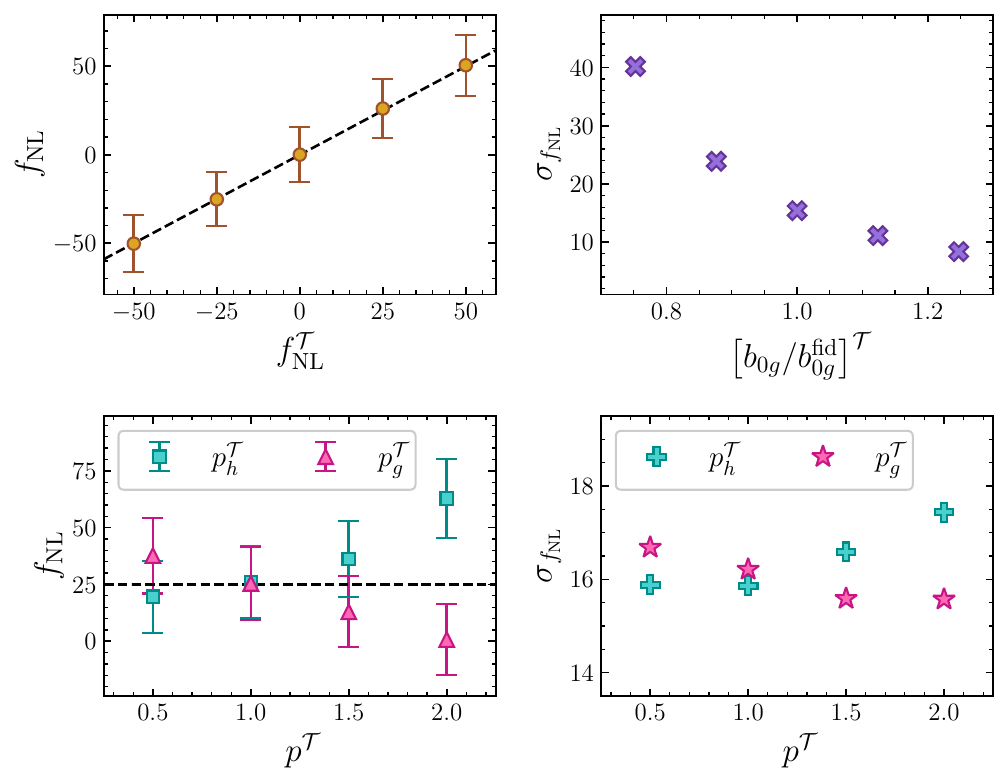}
\caption{Upper left: The measured values of $f_{\mathrm{NL}}$, with $1\sigma$ uncertainties, for SY5 toy model data where $f_{\mathrm{NL}}^{\mathcal{T}}$ is varied from $-50$ to $50$. The dashed line has a slope of 1 and intercept of 0. Upper right: The size of the uncertainties on the $f_{\mathrm{NL}}$ measurement, $\sigma_{f_{\mathrm{NL}}}$, for SY5 toy model data where $b_{0g}^{\mathcal{T}}$ is varied from $1.01$ to $1.68$. The uncertainties are given as a function of the ratio between the toy model linear bias, and the toy model fiducial linear bias, $[b_{0g} / b_{0g}^{\mathrm{fid}}]^{\mathcal{T}}$. Lower left: The measured values of $f_{\mathrm{NL}}$, with $1\sigma$ uncertainties, for SY5 toy model data where $f_{\mathrm{NL}}^{\mathcal{T}} = 25$ and the values of the $p$ parameters have been modified. Both $p_h^{\mathcal{T}}$ (cyan) and $p_g^{\mathcal{T}}$ (pink) were varied from $0.5$ to $2.0$. The priors on $p_h$ and $p_g$ have not been modified. The dashed line denotes $f_{\mathrm{NL}} = 25$. Lower right: The size of the uncertainties on the $f_{\mathrm{NL}}$ measurement, $\sigma_{f_{\mathrm{NL}}}$, for SY5 toy model data where $p$ is varied. This panel shows the magnitude of the uncertainties in the lower left panel.}
\label{fig:alt_poi_sys_tests}
\end{figure}

We use an SY5 toy model to produce 2pcfs and 3pcfs for two different cases - ($f_{\mathrm{NL}}^{\mathcal{T}} = 25$ and $b_{0g}^{\mathcal{T}} = 1.34$) and  ($f_{\mathrm{NL}}^{\mathcal{T}} = 0$, $b_{0g}^{\mathcal{T}} = 1.21$). We show the marginalized distributions over the POIs and nuisance parameters for these two additional cases in Fig.~\ref{fig:constraints_altPOI}. The POIs were correctly evaluated in both cases. We note that the uncertainty on $f_{\mathrm{NL}}$ increases with the decrease of $b_{0g}$. 

We perform additional systematic tests to understand the effects of the POIs and nuisance parameters on the uncertainties on  $f_{\mathrm{NL}}$. To ensure we retain the ability to measure $f_{\mathrm{NL}}$ in the extreme cases, we generate SY5 toy model data varying $f_{\mathrm{NL}}^{\mathcal{T}}$ from $-50$ to $50$. We show the measured  $f_{\mathrm{NL}}$ vs. input $f_{\mathrm{NL}}^{\mathcal{T}}$ in the upper left panel of Fig~\ref{fig:alt_poi_sys_tests}. The magnitude of the uncertainties are consistent across this range, and the measured $f_{\mathrm{NL}}$ is in agreement with the toy model value for all cases. While this test is performed using only toy model data, we also verify our model's ability to accurately constrain $f_{\mathrm{NL}}$ for a variety of non-zero PNG cases. This test is described in detail in Appendix~\ref{app:nonzero_png}.

As shown in Fig.~\ref{fig:constraints_altPOI}, a change in the linear galaxy bias relative to the fiducial simulation results in a change in the size of the $f_{\mathrm{NL}}$ uncertainties. To further explore this dependence, we generate a set of SY5 toy model data with $f_{\mathrm{NL}}^{\mathcal{T}} = 0$, and varied linear bias, $b_{0g}^{\mathcal{T}}$. In the upper right panel of Fig~\ref{fig:alt_poi_sys_tests}, we show the magnitude of the $f_{\mathrm{NL}}$ uncertainty as a function of the ratio between the toy model linear bias and the toy model fiducial linear bias. It is clear that tracers with a larger linear bias with respect to the fiducial model offer improved constraining power on $f_{\mathrm{NL}}$.

While we have chosen to center our priors on $p_h$ and $p_g$ on 1, this may not correspond to the true value for either tracer. To explore the sensitivity to the values of $p$, we generated toy model data with varied $p_h^{\mathcal{T}}$ and $p_g^{\mathcal{T}}$. Since this most dramatically affects the sensitivity when $f_{\mathrm{NL}}$ is nonzero, these toy model data sets were generated with $f_{\mathrm{NL}}^{\mathcal{T}} = 25$. We apply the same fitting procedure, with the priors centered  on 1. We show the measured $f_{\mathrm{NL}}$ and its uncertainty in the lower left and right panels of Fig~\ref{fig:alt_poi_sys_tests}. We retain the ability to measure $f_{\mathrm{NL}}$ to within $1\sigma$ of the true value when either $p_g$ or $p_h$ is varied by about 0.5. For a larger variation, our measurement of $f_{\mathrm{NL}}$ begins to deviate significantly from the true value. We also note that while this mismatch in $p$ between the true value and the prior affects the size of the uncertainties on $f_{\mathrm{NL}}$, this effect is small compared to the effect of varying the linear galaxy bias with respect to the fiducial simulation. This effect is also demonstrated by the shape of the $(f_{\mathrm{NL}}$, $p_h)$ and $(f_{\mathrm{NL}}$, $p_g)$ contours in Fig.~\ref{fig:constraints_altPOI}. If the priors on both $p$ parameters are accurate, we can expect larger uncertainties on $f_{\mathrm{NL}}$ in the case of higher $p_g$ and lower $p_h$. 

To summarize, the parameter that is most significantly correlated with $f_{\mathrm{NL}}$ is $b_{0g}$, specifically the measured value relative to the value in the fiducial simulation. The dependence on the values of $p_h$ and $p_g$ is considerably less pronounced.

\subsection{Advantages of large scales and higher correlation orders}
\label{subsec:sensitivity_gain}

\begin{figure}
\includegraphics[width=\columnwidth]{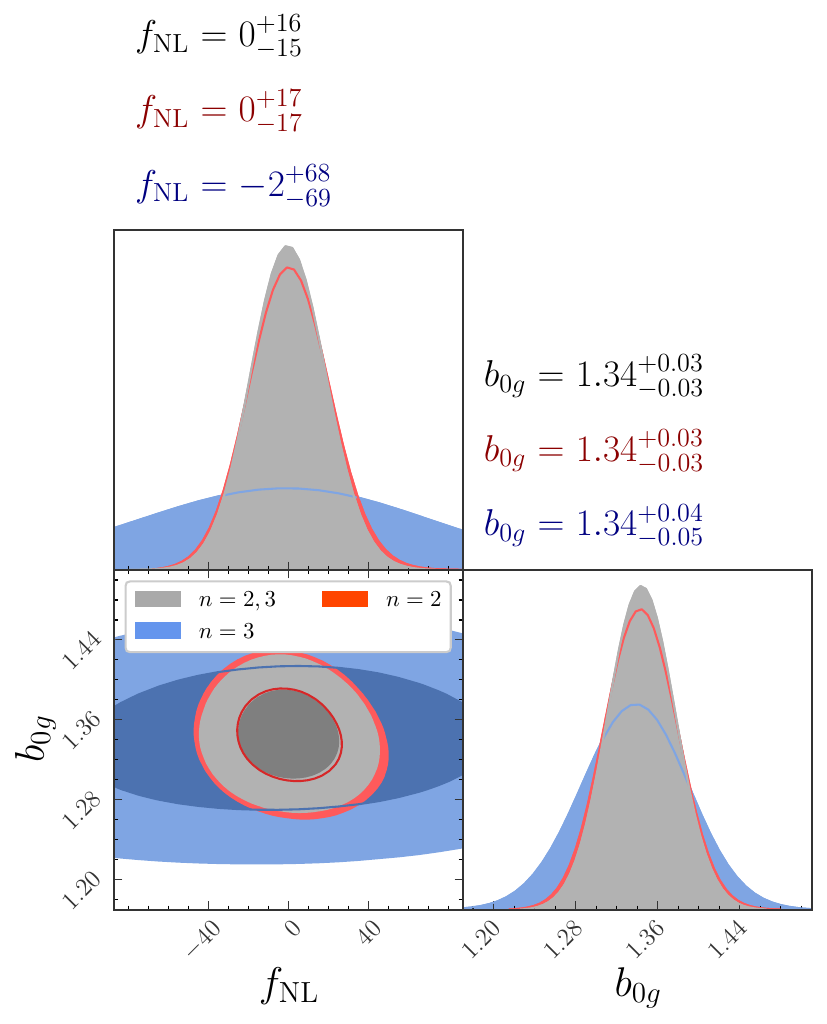}
\caption{Marginalized distributions of the POIs for SY5 toy model $n$pcfs, using test statistics constructed from the 2pcf and 3pcf combined (grey), 2pcf only (red) and 3pcf only (blue). The dark and light regions represent $1\sigma$ and $2\sigma$ contours, respectively. The most probable values and $1\sigma$ CL are labeled for each parameter above the respective panel.}
\label{fig:sens_corr_order}
\end{figure}

By combining the information contained in the 2pcf and 3pcf, we achieve greater sensitivity to the POIs than for either case alone. We demonstrate this by constructing the test statistic for SY5 toy model data using the 2pcf and 3pcf combined, as well as each separately. For all of these cases, we generate toy observations with $f_{\mathrm{NL}}^{\mathcal{T}} = 0$ and $b_{0g}^{\mathcal{T}} = 1.34$. We show the marginalized distributions of the POIs for this test in Fig.~\ref{fig:sens_corr_order}. By itself, the 3pcf monopole lacks the ability to meaningfully constrain the value of $f_{\mathrm{NL}}$. However, in combination with the information contained in the 2pcf, we improve the sensitivity.

\begin{table}
\begin{center}
\begin{threeparttable}
\begin{tabular}{lllll}
\toprule
2pcf scales & 3pcf scales & $N_{\mathrm{obs}}^{\mathrm{tot}}$ & $\sigma_{f_{\mathrm{NL}}}$ & $\sigma_{b_{0g}}$\\
\ [$h^{-1}$Mpc] & \ [$h^{-1}$Mpc] & \ & \ & \ \\
\midrule
50\ \mbox{---}\ 200 & 50\ \mbox{---}\ 200 & 152 & 22 & 0.029 \\
50\ \mbox{---}\ 260 & 50\ \mbox{---}\ 200 & 158 & 19 & 0.030 \\
50\ \mbox{---}\ 320 & 50\ \mbox{---}\ 200 & 164 & 18  & 0.029 \\
50\ \mbox{---}\ 380 & 50\ \mbox{---}\ 200 & 170 & 16  & 0.029 \\
\bottomrule
\end{tabular}
\caption{$\sigma_{f_{\mathrm{NL}}}$ and  $\sigma_{b_{0g}}$, the uncertainties on $f_{\mathrm{NL}}$ and $b_{0g}$ respectively, extracted using the SY5 toy model with $f_{\mathrm{NL}}^{\mathcal{T}} = 0$ and $b_{0g}^{\mathcal{T}} = 1.34$ by varying $s_{\mathrm{max}}$ of the 2pcf. $N_{\mathrm{obs}}^{\mathrm{tot}}$ is the total length of the combined 2pcf and 3pcf observable.}
\vspace{-0.2cm}
\label{tab:scale_test}
\end{threeparttable}
\end{center}
\end{table}

\begin{figure}
\includegraphics[width=\columnwidth]{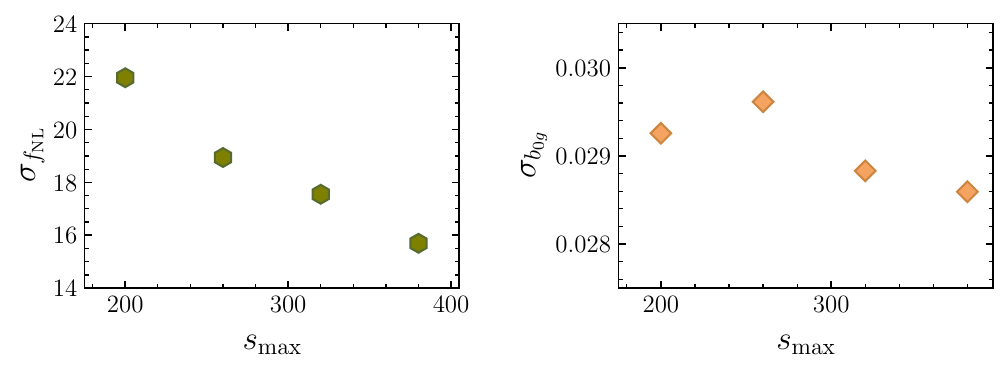}
\caption{$\sigma_{f_{\mathrm{NL}}}$ and $\sigma_{b_{1g}}$, the uncertainties on $f_{\mathrm{NL}}$ and $b_{0g}$ respectively, extracted from SY5 toy model data with $f_{\mathrm{NL}}^{\mathcal{T}} = 0$ and $b_{0g}^{\mathcal{T}} = 1.34$ as a function of a varying $s_{\mathrm{max}}$ in the vector of 2pcf observations.}
\label{fig:scale_test_plot}
\end{figure}

Given that most of the sensitivity to $f_{\mathrm{NL}}$ is extracted from the 2pcf, we may also wonder how much of the constraining power of the 2pcf comes from large scales. As cosmological surveys increase in volume, measurements of $n$pcfs at these scales will improve in their precision. We can demonstrate how galaxy clustering at large scales improves our own sensitivity to the POIs by constructing the test statistic using a limited range of scales. We generate SY5 toy model $n$pcfs, again corresponding to $f_{\mathrm{NL}}^{\mathcal{T}} = 0$ and $b_{0g}^{\mathcal{T}} = 1.34$, but vary the maximum clustering scale $s_{\mathrm{max}}$ of the 2pcf. We keep the separation between bins, $\Delta s = 10$ $h^{-1}$Mpc, and generate 2pcfs from $s = 50\ h^{-1}\mathrm{Mpc}\ \mbox{---}\ s_{\mathrm{max}}$. The scales used to generate the 3pcf for these test are held constant ($s = 50\ h^{-1}\mathrm{Mpc}\ \mbox{---}\ 200\ h^{-1}\mathrm{Mpc}$). The procedure is repeated for $s_{\mathrm{max}} = 200, 260, 320, 380$ $h^{-1}$Mpc.

We show the resulting $1\sigma$ uncertainties on the POIs in Table~\ref{tab:scale_test} and Fig.~\ref{fig:scale_test_plot}. In the table, we round the uncertainties on $b_{0g}$ to an additional decimal to show the minimal effect. Increasing the scales of the 2pcf does little to further constrain the linear bias. The sensitivity is nearly scale independent. The ability to measure $f_{\mathrm{NL}}$, however, depends strongly on $s_{\mathrm{max}}$. As we include larger clustering scales in the vector of 2pcf observations, we reduce the uncertainties on $f_{\mathrm{NL}}$. The diminishing nature of this effect is due to increasing cosmic variance in the $n$pcfs for surveys of the size used in this study. Thus, a larger survey (or a sample of LRGs combined with other tracers) occupying larger volume would offer even greater constraining power. Beyond the scales chosen for this analysis, however, we would likely need to include higher order $f_{\mathrm{NL}}$ terms.

\begin{figure}
\includegraphics[width=\columnwidth]{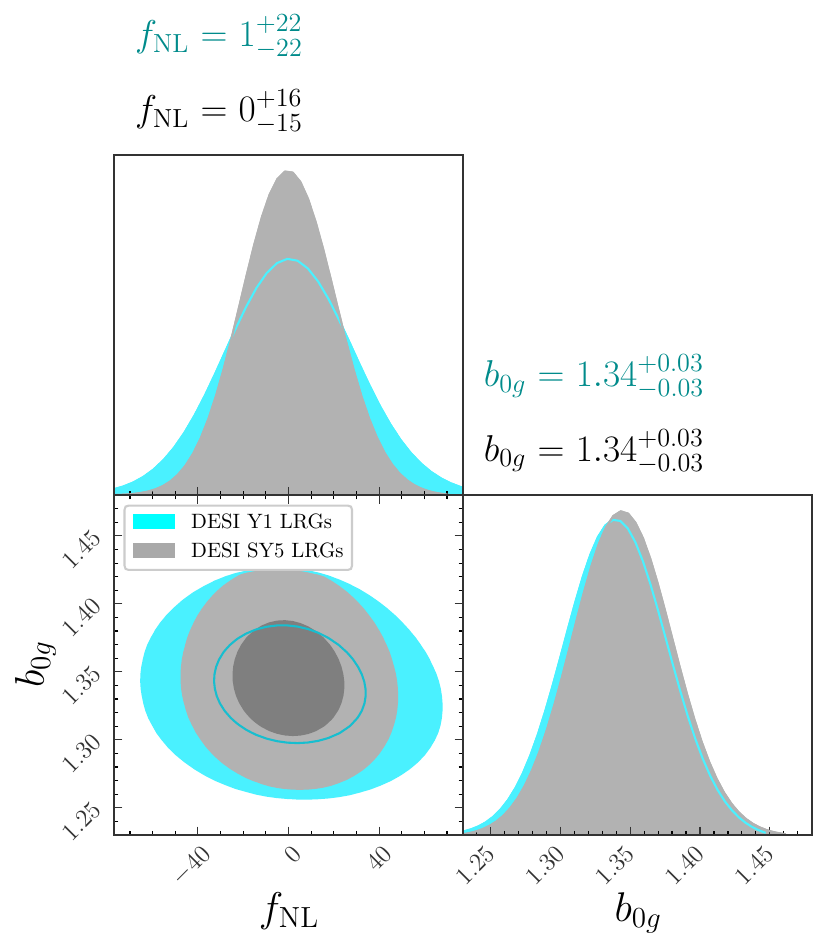}
\caption{Marginalized distributions of the POIs for toy model $n$pcfs corresponding to DESI SY5 LRGs (grey) and DESI Y1 LRGs (cyan). The dark and light regions represent $1\sigma$ and $2\sigma$ contours, respectively. The most probable values and $1\sigma$ CL are labeled for each parameter above the respective panel.}
\label{fig:sens_y1y5}
\end{figure}

\subsection{DESI Y1 LRG forecast}
\label{y1_forecast}

The DESI Y1 LRGs will cover a smaller volume than the full sample, as the survey is not yet complete. Thus, $n$pcfs corresponding to the ensemble of Y1 EZ mocks, whose coverage is shown in Fig.~\ref{fig:desi_window}, have larger variance. To forecast the sensitivity of our model to $f_{\mathrm{NL}}$ in the Y1 LRG survey, we repeat the procedure with the Y1 EZ mocks defining the fiducial model. We generate 2pcf and 3pcf Y1 toy model data using the covariance of this ensemble. In Fig.~\ref{fig:sens_y1y5}, we show the marginalized distributions of the POIs for the Y1 and SY5 toy model data with $f_{\mathrm{NL}}^{\mathcal{T}} = 0$. The effective redshift of the Y1 sample differs from that of the SY5 simulations. In the Y1 sample, we derive the Gaussian prior on $b_{0t}$ using $z_g = 0.73$. For the normalized linear bias, we find good agreement between the priors for the SY5 and Y1 samples, which informs the toy model value, $b_{0g}^{\mathcal{T}} = 1.34$ for both. We compare the projected precision for each, finding that for the Y1 LRGs, we can expect $\sigma_{f_{\mathrm{NL}}} \approx 22$ compared to $\sigma_{f_{\mathrm{NL}}} \approx 16$ for the SY5 toy model data. 
It is reasonable to assume that this sensitivity will continue to improve with larger volumes covered by future galaxy surveys.



\section{Discussion}
\label{sec:discussion}

Any measurement of local PNG from LSS would benefit from a robust confirmation, using two-point and three-point statistics in configuration and Fourier space. We have provided a method to achieve that. 

In this study, we limited the cases considered to the 2pcf and 3pcf monopole term for comparison to existing Fourier space studies.  {\tt ConKer} could be employed to measure $n$pcfs for higher correlation orders, which  may be even more sensitive to PNG. However, this approach  presents several difficulties. First, higher orders of correlation introduce additional PNG biases that may have non-trivial values. This would require a careful reconsideration of the second term in Eq.~\ref{eq:xi_mod_full}. The parameter $r_{\phi}$, which characterizes the ratio of PNG biases between the DM halos and the tracers, may not be sufficient. Furthermore, correlations of order $n>3$ may need to be parsed into their connected and disconnected terms \citep{philcox2021encore1}, which may introduce additional systematics in the case of non-zero PNG.

To apply the proposed method to observed LRGs, we also need to account for survey systematics, which can induce a scale-dependent deviation in the galaxy clustering signal. Imaging systematics from the DESI Legacy Imaging Surveys \citep{2017PASP..129f4101Z,2019AJ....157..168D,dr9} and survey completeness \citep{2023arXiv230606309S,expcalc,2023AJ....165...50M} are typically mitigated by weights, which should be introduced into the model as additional nuisance parameters and marginalized. Spurious effects of the spectroscopic pipeline \citep{2023AJ....165..144G,redrock2023,2023arXiv230510426B} will also be mitigated with redshift failure weights. If the analysis considers scales below $s = 50\ h^{-1}\mathrm{Mpc}$, we may also introduce nuisance parameters corresponding to variations in the Halo Occupancy Distribution (HOD). In addition, smaller scales are subject to variation arising from the process of assigning DESI fibers to targets \citep{fba}. These extensions of our model are the focus of ongoing research, targeted for a separate publication.

Finally, we note that while this study has demonstrated the application of our method to DESI LRGs, the model is applicable to an arbitrary matter tracer. All that is required is an ensemble of fiducial simulations for that tracer to evaluate the covariance matrix, and the expected clustering signal in the absence of PNG (like the EZ mocks), and a choice of the parameter, $p$. These simulations are generally less computationally expensive than full N-body simulations, and often accompany LSS data catalog releases. Therefore, this strategy is applicable to other DESI matter tracers, or those observed in future surveys like Euclid.



\section{Conclusions}
\label{sec:conclusions}

In summary, we developed a method by which local PNG, parameterized by $f_{\mathrm{NL}}$, is extracted from configuration space correlation functions. Using the 2pcf and 3pcf of FastPM-L3 halo simulations to characterize the scale dependent bias, we have estimated the sensitivity of DESI LRGs to this signal. We find competitive constraints on local PNG using toy model data, and confirm the robustness of this result using independent, high fidelity simulations. With an appropriate treatment of observational systematics, we believe this framework is applicable to current and future LSS surveys like DESI or Euclid, and will provide additional insight into the mechanisms of inflation.



\section*{Acknowledgements}

The authors would like to thank Z. Slepian, J. Hou, H. Gil-Mar\'{i}n, M. Wang, and L. Samushia for their interest and insightful comments, and S. BenZvi, K. Douglass, and J. Bermejo Climent for useful discussions. 


Z. Brown and R. Demina acknowledge support from the U.S. Department of Energy under the grant DE-SC0008475.0. 

This material is based upon work supported by the U.S. Department of Energy (DOE), Office of Science, Office of High-Energy Physics, under Contract No. DE–AC02–05CH11231, and by the National Energy Research Scientific Computing Center, a DOE Office of Science User Facility under the same contract. 

Additional support for DESI was provided by the U.S. National Science Foundation (NSF), Division of Astronomical Sciences under Contract No. AST-0950945 to the NSF’s National Optical-Infrared Astronomy Research Laboratory; the Science and Technology Facilities Council of the United Kingdom; the Gordon and Betty Moore Foundation; the Heising-Simons Foundation; the French Alternative Energies and Atomic Energy Commission (CEA); the National Council of Science and Technology of Mexico (CONACYT); the Ministry of Science and Innovation of Spain (MICINN), and by the DESI Member Institutions: \url{https://www.desi.lbl.gov/collaborating-institutions}. 

Any opinions, findings, and conclusions or recommendations expressed in this material are those of the author(s) and do not necessarily reflect the views of the U. S. National Science Foundation, the U. S. Department of Energy, or any of the listed funding agencies. 

The authors are honored to be permitted to conduct scientific research on Iolkam Du’ag (Kitt Peak), a mountain with particular significance to the Tohono O’odham Nation.
%

\section*{Data Availability}


All simulations of LRGs included in this study (EZ mocks and AbacusSummit simulations) will be made available in upcoming data releases from the Dark Energy Spectroscopic Instrument. Please contact the authors for access to the collection of Fast-PM simulations. The data shown in this paper's figures are available in machine-readable format at \url{https://zenodo.org/records/10794474}.


\bibliographystyle{mnras}
\bibliography{bib} 



\appendix

\section{Author Affiliations}
\label{app:affiliations}

{\it 
$^{1}$ Department of Physics and Astronomy, University of Rochester, 500 Joseph C. Wilson Boulevard, Rochester, NY 14627, USA \\
$^{2}$ Instituto de F\'{\i}sica Te\'{o}rica (IFT) UAM/CSIC, Universidad Aut\'{o}noma de Madrid, Cantoblanco, E-28049, Madrid, Spain \\
$^{3}$ Institut de F\'{i}sica d'Altes Energies (IFAE), The Barcelona Institute of Science and Technology, Campus UAB, 08193 Bellaterra Barcelona, Spain \\
$^{4}$ Lawrence Berkeley National Laboratory, 1 Cyclotron Road, Berkeley, CA 94720, USA \\
$^{5}$ Centro de Investigaci\'{o}n Avanzada en F\'{\i}sica Fundamental (CIAFF), Facultad de Ciencias, Universidad Aut\'{o}noma de Madrid, ES-28049 Madrid, Spain \\
$^{6}$ Physics Dept., Boston University, 590 Commonwealth Avenue, Boston, MA 02215, USA \\
$^{7}$ NSF's NOIRLab, 950 N. Cherry Ave., Tucson, AZ 85719, USA \\
$^{8}$ Department of Physics \& Astronomy, University College London, Gower Street, London, WC1E 6BT, UK \\
$^{9}$ Institute for Computational Cosmology, Department of Physics, Durham University, South Road, Durham DH1 3LE, UK \\
$^{10}$ Instituto de F\'{\i}sica, Universidad Nacional Aut\'{o}noma de M\'{e}xico,  Cd. de M\'{e}xico  C.P. 04510,  M\'{e}xico \\
$^{11}$ Department of Physics \& Astronomy and Pittsburgh Particle Physics, Astrophysics, and Cosmology Center (PITT PACC), University of Pittsburgh, 3941 O'Hara Street, Pittsburgh, PA 15260, USA \\
$^{12}$ Kavli Institute for Particle Astrophysics and Cosmology, Stanford University, Menlo Park, CA 94305, USA \\
$^{13}$ SLAC National Accelerator Laboratory, Menlo Park, CA 94305, USA \\ 
$^{14}$ Departamento de F\'isica, Universidad de los Andes, Cra. 1 No. 18A-10, Edificio Ip, CP 111711, Bogot\'a, Colombia \\
$^{15}$ Observatorio Astron\'omico, Universidad de los Andes, Cra. 1 No. 18A-10, Edificio H, CP 111711 Bogot\'a, Colombia \\
$^{16}$ Institut d'Estudis Espacials de Catalunya (IEEC), 08034 Barcelona, Spain \\
$^{17}$ Institute of Cosmology \& Gravitation, University of Portsmouth, Dennis Sciama Building, Portsmouth, PO1 3FX, UK \\
$^{18}$ Institute of Space Sciences, ICE-CSIC, Campus UAB, Carrer de Can Magrans s/n, 08913 Bellaterra, Barcelona, Spain \\
$^{19}$ Center for Cosmology and AstroParticle Physics, The Ohio State University, 191 West Woodruff Avenue, Columbus, OH 43210, USA \\
$^{20}$ Department of Physics, The Ohio State University, 191 West Woodruff Avenue, Columbus, OH 43210, USA \\
$^{21}$ The Ohio State University, Columbus, 43210 OH, USA \\
$^{22}$ School of Mathematics and Physics, University of Queensland, 4072, Australia \\
$^{23}$ Department of Physics, Southern Methodist University, 3215 Daniel Avenue, Dallas, TX 75275, USA \\
$^{24}$ Sorbonne Universit\'{e}, CNRS/IN2P3, Laboratoire de Physique Nucl\'{e}aire et de Hautes Energies (LPNHE), FR-75005 Paris, France \\
$^{25}$ Departament de F\'{i}sica, Serra H\'{u}nter, Universitat Aut\`{o}noma de Barcelona, 08193 Bellaterra (Barcelona), Spain \\
$^{26}$ Instituci\'{o} Catalana de Recerca i Estudis Avan\c{c}ats, Passeig de Llu\'{\i}s Companys, 23, 08010 Barcelona, Spain \\
$^{27}$ Department of Physics and Astronomy, University of Sussex, Brighton BN1 9QH, U.K \\
$^{28}$ Department of Physics \& Astronomy, University  of Wyoming, 1000 E. University, Dept.~3905, Laramie, WY 82071, USA \\
$^{29}$ National Astronomical Observatories, Chinese Academy of Sciences, A20 Datun Rd., Chaoyang District, Beijing, 100012, P.R. China \\
$^{30}$ Departamento de F\'{i}sica, Universidad de Guanajuato - DCI, C.P. 37150, Leon, Guanajuato, M\'{e}xico \\
Instituto Avanzado de Cosmolog\'{\i}a A.~C., San Marcos 11 - Atenas 202. \\ $^{31}$ Magdalena Contreras, 10720. Ciudad de M\'{e}xico, M\'{e}xico
$^{32}$ IRFU, CEA, Universit\'{e} Paris-Saclay, F-91191 Gif-sur-Yvette, France \\
$^{33}$ Space Sciences Laboratory, University of California, Berkeley, 7 Gauss Way, Berkeley, CA  94720, USA \\
$^{34}$ University of California, Berkeley, 110 Sproul Hall \#5800 Berkeley, CA 94720, USA \\
$^{35}$ Department of Physics, Kansas State University, 116 Cardwell Hall, Manhattan, KS 66506, USA \\
$^{36}$ Department of Physics and Astronomy, Sejong University, Seoul, 143-747, Korea \\
$^{37}$ CIEMAT, Avenida Complutense 40, E-28040 Madrid, Spain \\
$^{38}$ Space Telescope Science Institute, 3700 San Martin Drive, Baltimore, MD 21218, USA \\
$^{39}$ Department of Physics, University of Michigan, Ann Arbor, MI 48109, USA \\
$^{40}$ University of Michigan, Ann Arbor, MI 48109, USA \\
}


\section{Assumption of Linearity}
\label{app:linearity}

The primary assumption of the statistical model described in Section~\ref{sec:method} is that the deviations in the 2pcf and 3pcf from the Gaussian case are linear in $f_{\mathrm{NL}}$. However, the theoretical 2pcf in the presence of PNG contains a term which scales as $f_{\mathrm{NL}}^2$. Similarly, the 3pcf with PNG contains a term which scales as $f_{\mathrm{NL}}^3$. However, the smallness of $\alpha^{-1}$ allows us to claim that for the scales considered in this analysis, the linear term is dominant.

We verify this assumption using an additional set of larger volume FastPM simulations with PNG \citep{chaussidoninprep}. Using the same FastPM method as the simulations described in Section~\ref{subsec:sim_details}, we generate four simulations in $(5.52\ h^{-1}\mathrm{Gpc})^3$ boxes evolved to $z=1.5$. These larger simulations are run with $f_{\mathrm{NL}} = -25, 0, 12, 25$. We again create catalogs by selecting halos with $M > 1.15\times10^{13}\ h^{-1}M_{\odot}$.

\begin{figure}
\includegraphics[width=\columnwidth]{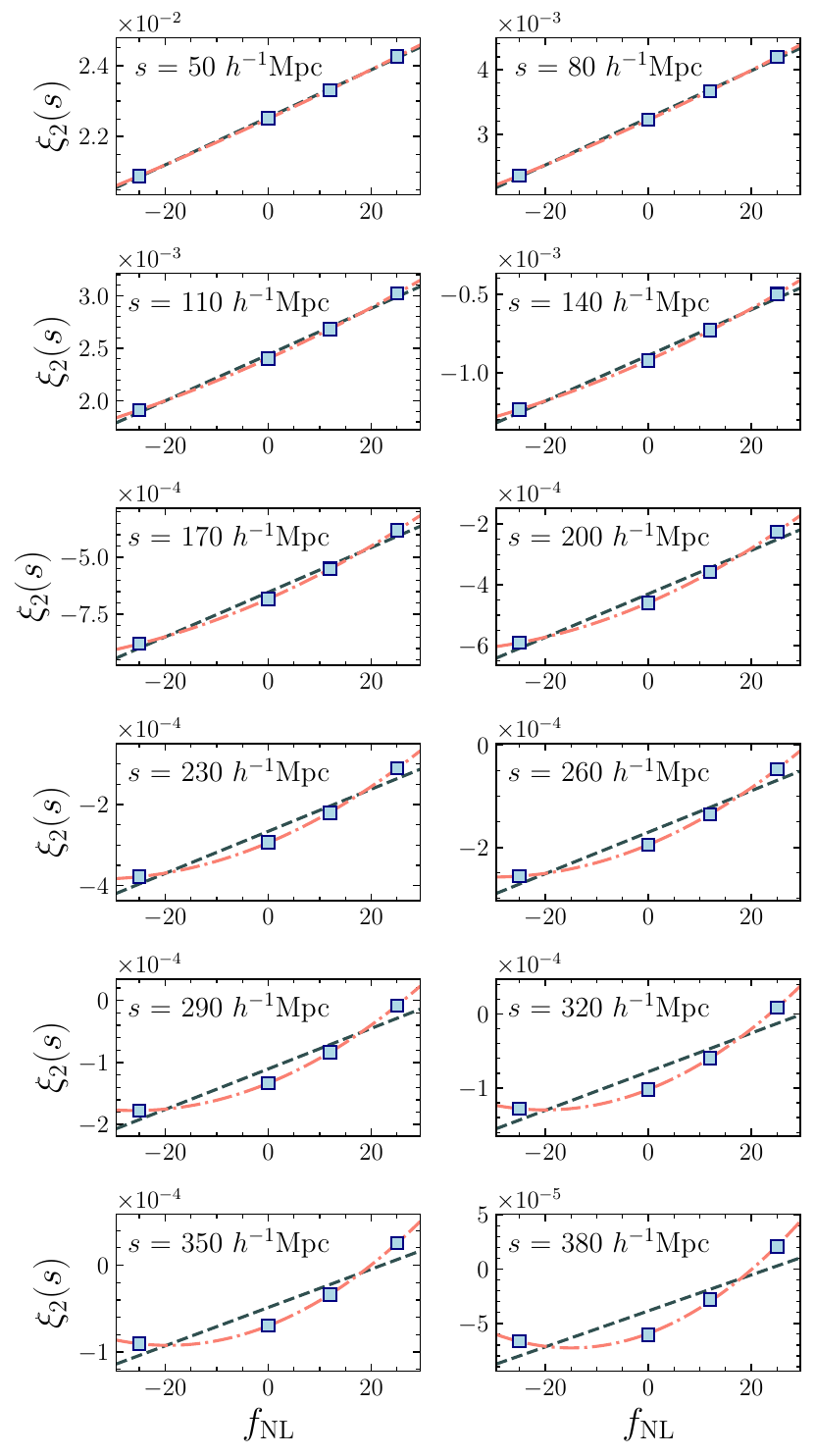}
\caption{The 2pcf of the FastPM-L5.52 simulations in selected bins, shown as a function of $f_{\mathrm{NL}}$ (blue markers). The grey dashed line denotes a linear fit while the red dot-dash line denotes a quadratic fit.}
\label{fig:lin_test_2pcf}
\end{figure}

For each one, we measure the 2pcf and diagonal 3pcf using {\tt ConKer} at the same scales as the ones used in our model. Fig.~\ref{fig:lin_test_2pcf} shows the magnitude of the 2pcf in selected bins plotted as a function of $f_{\mathrm{NL}}$. For each selected bin, we perform both a linear and quadratic fit. At the lowest scales, the linear assumption is extremely well motivated, and we observe almost no difference between the linear and quadratic fits. As the scale increases we begin to see deviations from linearity. The decision to truncate our model at $s = 380\  h^{-1}\mathrm{Mpc}$ is motivated by this breakdown of linearity. A more conservative upper limit at a lower scale would better preserve the linearity of the 2pcf, but increase the uncertainties on $f_{\mathrm{NL}}$. In order to extend the model beyond these scales, we would likely need to model both the linear and quadratic $f_{\mathrm{NL}}$ terms, which would significantly complicate the procedure.

\begin{figure}
\includegraphics[width=\columnwidth]{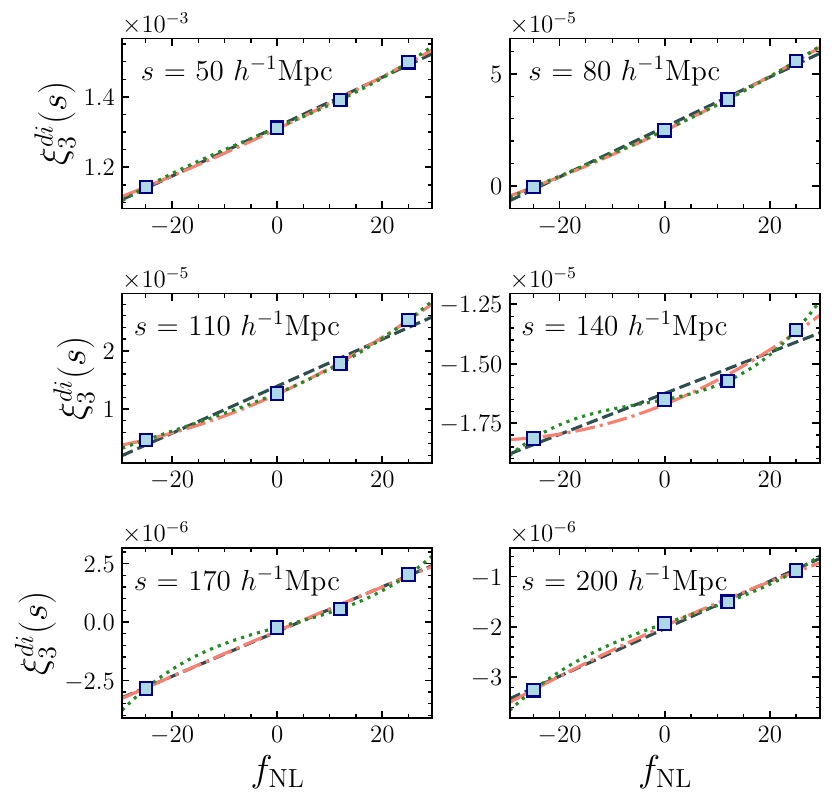}
\caption{The diagonal 3pcf of the FastPM-L5.52 simulations in selected bins, shown as a function of $f_{\mathrm{NL}}$ (blue markers). The grey dashed line denotes a linear fit, the red dot-dash line denotes a quadratic fit, and the green dotted line denotes a cubic fit.}
\label{fig:lin_test_3pcf}
\end{figure}

We perform a similar test for the diagonal 3pcf, which is shown in Fig.~\ref{fig:lin_test_3pcf}. Here, we also include a cubic fit as the 3pcf contains a term which scales as $f_{\mathrm{NL}}^3$. For the 3pcf, we truncate the model at $s = 200\  h^{-1}\mathrm{Mpc}$. Fig.~\ref{fig:lin_test_3pcf} shows that below this scale, the assumption of linearity in $f_{\mathrm{NL}}$ is justified. For the 3pcf when $s_1$ is not equal to $s_2$, we exclude any bins where either $s_1$ or $s_2$ is larger than $s = 200\  h^{-1}\mathrm{Mpc}$.

These tests, performed using the FastPM-L5.52 simulations, motivate the choice of scales in our statistical model. For the 2pcf when $50\  h^{-1}\mathrm{Mpc} < s < 380\  h^{-1}\mathrm{Mpc}$ and for the 3pcf when $50\  h^{-1}\mathrm{Mpc} < s_1, s_2 < 200\  h^{-1}\mathrm{Mpc}$, we are confident that changes in the clustering signal due to PNG are linear in $f_{\mathrm{NL}}$.

\section{Non-zero PNG validation}
\label{app:nonzero_png}

In Section~\ref{subsec:Results}, we demonstrate the ability of our method to correctly measure $f_{\mathrm{NL}}$ in independent  simulations without PNG ($f_{\mathrm{NL}} = 0$). To validate our method, we must also verify that it can correctly measure $f_{\mathrm{NL}}$ in the presence of PNG. The AbacusSummit simulations are unsuitable for this test, as they have only been generated with Gaussian initial conditions. However, we can use the FastPM-L5.52 simulations described in Appendix~\ref{app:linearity} for such a test.

We begin by constructing the statistical model using the FastPM-L3 simulations to interpolate the values of $A_n$. For the test case, we use each of the FastPM-L5.52 simulations, where $f_{\mathrm{NL}} = -25, 0, 12, 25$. The fiducial $n$pcfs, or $\xi_n^{\mathrm{fid}}$, are derived from the $f_{\mathrm{NL}} = 0$ case of these larger simulations. The covariance matrix is estimated by scaling the covariance of the FastPM-L3 simulations with $f_{\mathrm{NL}} = 0$ according to the difference in volume. This is required because the computational burden of generating sufficiently many FastPM-L5.52 simulations to evaluate the covariance is too great. 

\begin{figure}
\includegraphics[width=\columnwidth]{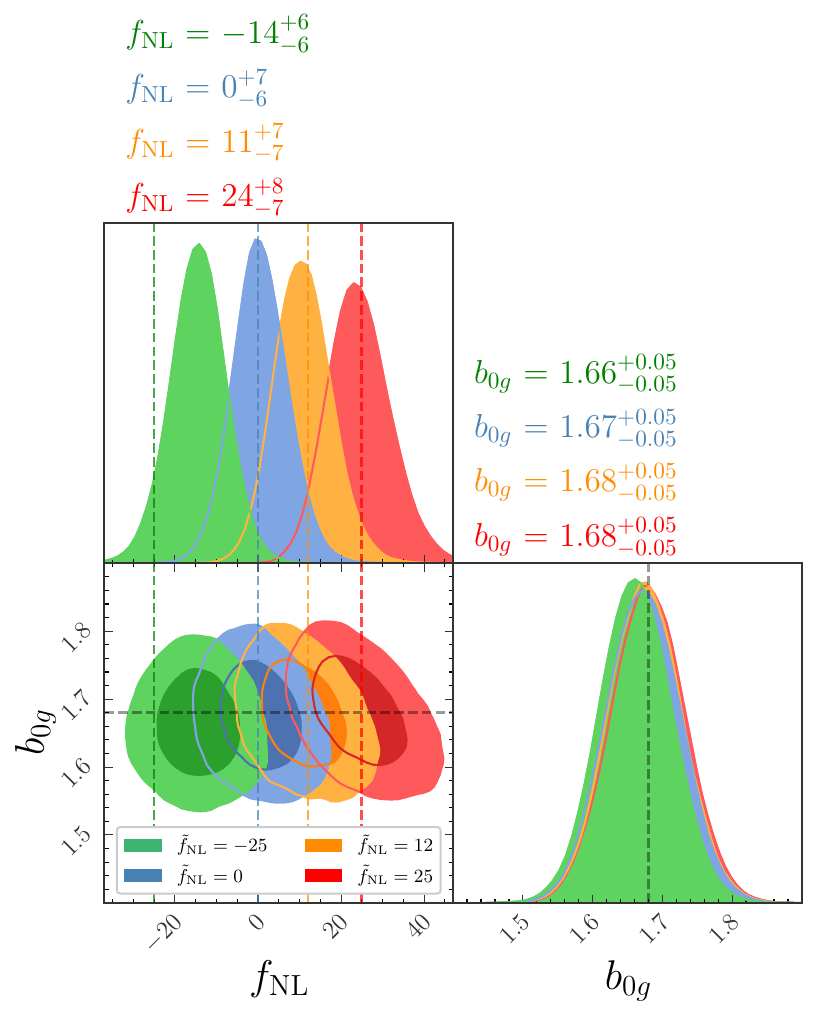}
\caption{Marginalized distributions of the POIs for each of the FastPM-L5.52 simulations (green, blue, orange, and red corresponding to $f_{\mathrm{NL}} = -25, 0, 12, 25$ respectively). The dark and light regions represent $1\sigma$ and $2\sigma$ contours, respectively. The most probable values and $1\sigma$ CL are labeled for each parameter above the respective panel. The dashed lines indicate the true values, which are also given as $\tilde{f}_{\mathrm{NL}}$ in the legend, where the tilde indicates truth.}
\label{fig:fastpm_xxxl_validation}
\end{figure}

In Fig.~\ref{fig:fastpm_xxxl_validation}, we show the marginalized distributions of the POIs for each case. For the zero and positive $f_{\mathrm{NL}}$ cases, the true value falls within the $1\sigma$ contours. For the case where $f_{\mathrm{NL}} = -25$, we slightly underestimate the value at between $1\sigma$ and $2\sigma$. We do, however, expect a slight deviation between the true and measured values, due in part to the halo finding procedure employed in the simulations. The halos in these FastPM simulations are determined from a Friends-of-Friends algorithm \citep{roy2014pfof}, and \cite{chaussidoninprep} showed that it reduces the amplitude of the $f_{\mathrm{NL}}$ signal up to 30\%, which was also noted in \cite{biagetti2017verifying}. The values of $f_{\mathrm{NL}}$ that we measure are compatible with the ones measured in \cite{chaussidoninprep}. We may also expect deviation between the true and measures values due to our linear parameterization. From Fig.~\ref{fig:lin_test_2pcf}, we see that points corresponding to positive values of $f_{\mathrm{NL}}$ are still in the linear regime, while points corresponding to negative $f_{\mathrm{NL}}$ deviate from linearity. This is because the positive 0$^{\mathrm{th}}$ and negative 1$^{\mathrm{st}}$ order terms cancel, making the 2$^{\mathrm{nd}}$ order term more significant. Since the parameterization was derived from mocks with positive $f_{\mathrm{NL}}$, it is not surprising that the results of the fit show better agreement for positive $f_{\mathrm{NL}}$, while deviating more significantly from the true value for the negative $f_{\mathrm{NL}}$ cases.

This test demonstrates the ability of our model to accurately constrain $f_{\mathrm{NL}}$ for an independent set of simulations at a different effective redshift than those used to generate the values of $A_n$. From this, we remain confident in the ability to apply these techniques to the upcoming DESI data releases.

\bsp	
\label{lastpage}
\end{document}